\documentclass[twocolumn,showpacs,preprintnumbers,amsmath,amssymb]{revtex4}

\usepackage{graphicx}
\usepackage{dcolumn}
\usepackage{bm}

\begin{document}

\preprint{ }

\title{Binding Energy of Charged Excitons in ZnSe-based Quantum Wells}

\author{G.~V.~Astakhov}
\author{D.~R.~Yakovlev}
\affiliation{ Physikalisches Institut der Universit\"{a}t
W\"{u}rzburg, 97074 W\"{u}rzburg, Germany and \\
A.F.Ioffe Physico-Technical Institute, Russian Academy of
Sciences, 194017, St.Petersburg, Russia }

\author{V.~P.~Kochereshko}
\affiliation{ A.F.Ioffe Physico-Technical Institute, Russian
Academy of Sciences, 194017, St.Petersburg, Russia }

\author{W.~Ossau}
\affiliation{ Physikalisches Institut der Universit\"{a}t
W\"{u}rzburg, 97074 W\"{u}rzburg, Germany }

\author{J.~Puls}
\author{F.~Henneberger}
\affiliation{ Humboldt-Universit\"{a}t zu Berlin, Institut f\"{u}r
Physik, 10115 Berlin, Germany }

\author{S.~A.~Crooker}
\author{Q.~McCulloch}
\affiliation{ National High Magnetic Field Laboratory, Los Alamos,
New Mexico 87545, USA }

\author{D.~Wolverson}
\affiliation{ University of Bath, BA2 7AY Bath, United Kingdom }

\author{N.~A.~Gippius}
\affiliation{ General Physics Institute, Russian Academy of
Sciences, 117333 Moscow, Russia }

\author{W.~Faschinger}
\affiliation{ Physikalisches Institut der Universit\"{a}t
W\"{u}rzburg, 97074 W\"{u}rzburg, Germany }

\author{A.~Waag}
\affiliation{ Abteilung Halbleiterphysik, Universit\"{a}t Ulm,
89081 Ulm, Germany }

\date{\today}

\begin{abstract}
Excitons and charged excitons (trions) are investigated in
ZnSe-based quantum well structures with (Zn,Be,Mg)Se and
(Zn,Mg)(S,Se) barriers by means of magneto-optical spectroscopy.
Binding energies of negatively- ($X^{-}$) and positively ($X^{ +
}$) charged excitons are measured as functions of quantum well
width, free carrier density and in external magnetic fields up to
47~T. The binding energy of $X^{ - }$ shows a strong increase from
1.4 to 8.9~meV with decreasing quantum well width from 190 to
29~{\AA}. The binding energies of $X^{ + }$ are about 25{\%}
smaller than the $X^{ - }$ binding energy in the same structures.
The magnetic field behavior of $X^{ - }$ and $X^{ + }$ binding
energies differ qualitatively. With growing magnetic field
strength, $X^{ - }$ increases its binding energy by 35-150{\%},
while for $X^{ + }$ it decreases by 25{\%}. Zeeman spin splittings
and oscillator strengths of excitons and trions are measured and
discussed.
\end{abstract}


\maketitle


\section{\label{sec1} INTRODUCTION }

Charged excitons (or trions) are exciton complexes consisting of
three particles. Two electrons and one hole form a negatively
charged exciton $X^{ - }$. Two holes and one electron can be
organized in a positively charged exciton $X^{ + }$. Trion
complexes in bulk semiconductors, i.e. in three dimensions, are
fragile, but become stable in low-dimensional systems. That is why
the theoretical prediction of Lampert from 1958 \cite{ref40} was
followed by a confident experimental observation of trions only in
1993 for the quasi-two dimensional electronic system in
CdTe/(Cd,Zn)Te quantum wells (QW's) \cite{ref41}. Since,
positively- and negatively charged excitons have been studied
experimentally in III-V heterostructures based on GaAs and in
II-VI quantum well structures based on CdTe, (Cd,Mn)Te, ZnSe and
(Zn,Mn)Se (see e.g. \cite{ref42, ref47, ref11, ref37} and
references therein).

II-VI semiconductors are very suitable for the trion studies due
to their strong Coulombic interaction compared with III-V
materials. E.g. exciton binding energies (exciton Rydberg) in
GaAs, CdTe and ZnSe are 4.2, 10 and 20~meV, respectively. Among
these materials ZnSe has the strongest Coulombic interaction.
However, after the first report of $X^{ - }$ observation in
Zn$_{0.9}$Cd$_{0.1}$Se/ZnSe QW's in 1994 \cite{ref30}, detailed
investigations were started from 1998 only, when the high-quality
ZnSe-based structures with binary quantum well layers were
fabricated \cite{ref10, ref34, ref35}. At present rather detailed
experimental information on trions in ZnSe QW's is available: (i)
negatively- and positively charged excitons were documented
\cite{ref11}; (ii) trions were reported for the light-hole
excitons \cite{ref10, ref11}; (iii) singlet- and triplet trion
states were studied in high magnetic fields \cite{ref32, ref16};
(iv) spin structure of trions and spin-dependent formation process
of trions were investigated \cite{ref13, ref32}; (v) recombination
dynamics in magnetic fields \cite{ref14} and coherent dynamics of
trions \cite{ref33} were studied; (vi) oscillator strength of
trion resonances was examined for different electron densities and
in magnetic fields \cite{ref12}. Theoretical results for this
material system are limited to a calculation of the trion binding
energy {\it vs} well width \cite{ref36} and its variation in high
magnetic fields \cite{ref16}. Agreement with experiment was rather
qualitative - one of the reasons for this is the uncertainty in
the parameters used for the calculations.

\begin{figure}
\includegraphics{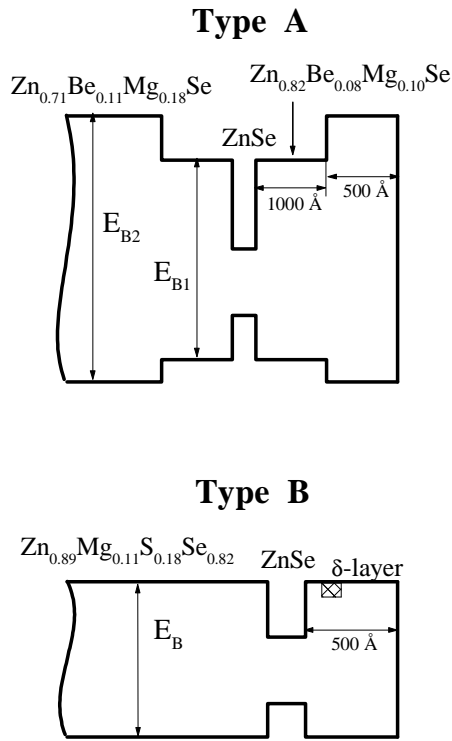}
\caption{\label{fig1} Schematics of the studied structures. Type~A
is ZnSe/Zn$_{0.82}$Be$_{0.08}$Mg$_{0.10}$Se single QW surrounded
by additional Zn$_{0.71}$Be$_{0.11}$Mg$_{0.18}$Se barrier. Type~B
is ZnSe/Zn$_{0.89}$Mg$_{0.11}$S$_{0.18}$Se$_{0.82}$ single QW with
modulation doping.}
\end{figure}

In this paper we present a detailed study of trion binding
energies in ZnSe-based structures as a function of quantum well
width and applied magnetic field. Parameters of exciton and trion
states were determined by means of magneto-optical experiments,
calculated on the base of variational approach, and evaluated from
the best fit of the experimental dependencies. The paper is
organized as follows: Sec.~\ref{sec2} details the structures,
while exciton parameters (measured and calculated) are discussed
in Sec.~\ref{sec3}. In Sec.~\ref{sec4} results on the binding
energies of trions are collected and discussed. Finally, in
Sec.~\ref{sec5} results of a modification of the singlet trion
state in high magnetic fields and with increasing carrier density
are presented.

In this paper we deal with positively- ($X^{ + }$) and negatively
($X^{ - }$ ) charged excitons. We will label them in this way when
the difference in the charge structure of trions is important. The
term ``trion'' ($T$) will be used as a general definition for both
positively- and negatively charged excitons.


\section{\label{sec2} EXPERIMENTALS}

{\it Quantum well structures}. ZnSe-based quantum well
heterostructures with a binary material of QW were grown by
molecular-beam epitaxy on (100)-oriented GaAs substrates. Studied
structures contain single quantum wells, with thickness varying
from 29 to 190~{\AA}. Schemes for the structure designs are
presented in Fig.~\ref{fig1}. Different barrier materials were
used, namely (Zn,Be,Mg)Se, (Zn,Mg)(S,Se) and (Zn,Be)Se, as shown
in Table~\ref{tab1}. To each barrier material the type of a
structure (A, B, C or D) has been assigned. Parameters of the
barrier materials were chosen with an aim to make them
lattice-matched to GaAs substrates, which allows growing QW's of a
very high structural quality.

\begin{table}
\caption{\label{tab1}Parameters of the barrier materials in
ZnSe-based QW's with a type-I band alignment. }
\begin{ruledtabular}
\begin{tabular}{ccccc}
 & &  Barrier & $\Delta E_g $ & ${\Delta E_C } \mathord{\left/ {\vphantom {{\Delta
E_C } {\Delta E_V }}} \right. \kern-\nulldelimiterspace} {\Delta
E_V }$\\

MATERIAL &  Type & band gap, & to ZnSe & to ZnSe \\
 &  & $E_g^b $ (eV) & (meV) &  \\
\hline
Zn$_{0.82}$Be$_{0.08}$Mg$_{0.10}$Se  & A & 3.06& 240 & 78/22 \\
first barrier  &  &  &  & \\
Zn$_{0.71}$Be$_{0.11}$Mg$_{0.18}$Se  & A & 3.21& 390 & 78/22 \\
second barrier  &  &  &  & \\
Zn$_{0.96}$Be$_{0.04}$Se & D & 2.89& 70 & 78/22 \\
\hline
Zn$_{0.89}$Mg$_{0.11}$S$_{0.18}$Se$_{0.82}$& B & 3.02& 200& 50/50 \\
Zn$_{0.95}$Mg$_{0.05}$S$_{0.09}$Se$_{0.91}$ & C & 2.92& 100& 50/50 \\

\end{tabular}
\end{ruledtabular}
\end{table}

Most of the structures used in this study were nominally undoped
(types A,C, and D). A background carrier density in them was tuned
by an additional above-barrier illumination. The range of tuning
depends on the QW width, allowing in the widest QW to vary
electron density from 5$\times $10$^{9}$ to 10$^{11}$~cm$^{ - 2}$.
Details of the illumination technique will be presented in
Sec.~\ref{sub4B}. Two structures of type B were modulation-doped
in the barrier layers. In the sample zq1038 free electrons in the
QW were provided by $n$-type doping with a 30-{\AA}-thick, Cl
doped layer (donor concentration of 5$\times $10$^{17}$~cm$^{ -
3})$ separated from the QW by a 100-{\AA}-thick spacer. The sample
zq1113 was $p$-type doped with nitrogen (RF plasma cell at a power
of 350~W and a background pressure of 5$\times $10$^{ - 6}$ Torr).
In this sample, symmetric doping was achieved by uniform doping of
barriers excluding 30-{\AA}-thick spacer layers. The concentration
of the two-dimensional hole gas (2DHG) in the QW of this sample is
about $n_h \approx $3$\times $10$^{10}$~cm$^{ - 2}$ and was
insensitive to additional illumination.

{\it Strain effect on the band gap}. ZnSe quantum well layers in
the studied structures experience compressive strain due to a
small difference in lattice constants of ZnSe and GaAs. This
results in an increase of the band gap, which is different for the
heavy-hole and light-hole states (for details see e.g.
Refs.~\onlinecite{ref2, ref3}). The calculated values of the band
gap increase give 2~meV for the heavy-hole states and 16~meV for
the light-hole states (required parameters of the elastic
stiffness constants $C_{11} $, $C_{12} $ and deformation
potentials $a$, $b$ are given in Table~\ref{tab2}).

{\it Band offsets}. (Zn,Be,Mg)Se or (Zn,Mg)(S,Se) barrier
materials differ by their band gap discontinuity to ZnSe quantum
well $\Delta E_g $ and its distribution between the conduction and
valence bands ${\Delta E_C } \mathord{\left/ {\vphantom {{\Delta
E_C } {\Delta E_V }}} \right. \kern-\nulldelimiterspace} {\Delta
E_V }$ (where $\Delta E_C + \Delta E_V = \Delta E_g )$. The band
gap discontinuity between the gap of ZnSe $E_g
(\mbox{ZnSe})$=2.82~eV (at $T$=1.6~K) and the barrier gap is
distributed between conduction and valence bands in proportion
${\Delta E_C } \mathord{\left/ {\vphantom {{\Delta E_C } {\Delta
E_V }}} \right. \kern-\nulldelimiterspace} {\Delta E_V }$=78/22
for ZnSe/(Zn,Be,Mg)Se heterointerface \cite{ref18}. For the
ZnSe/(Zn,Mg)(S,Se) heterosystem, different values can be found in
literature varying from 50/50 \cite{ref54, ref55} to 10/90
\cite{ref56}. We chose a 50/50 ratio for our calculation of the
structure parameters. We believe that very similar values for
trion binding energies measured in the both types of studied
structures (to be shown in Fig.~\ref{fig16}) justify a
considerable confinement for electrons and, respectively, approve
our choice of ${\Delta E_C } \mathord{\left/ {\vphantom {{\Delta
E_C } {\Delta E_V }}} \right. \kern-\nulldelimiterspace} {\Delta
E_V }$. In Sec.~\ref{sec3} we will use, for consideration of
exciton parameters, published experimental data for ZnSe-based
QW's with other barrier materials, namely, (Zn,Be)Se and
(Zn,Mg)(S,Se) of lower content. Parameters for the barrier
materials are collected in Table~\ref{tab1}.

\begin{table}
\caption{\label{tab2}Parameters of ZnSe and ZnSe-based QW's with
type-I band alignment. }
\begin{ruledtabular}
\begin{tabular}{cc}
``Unstrained'' band gap $E_g$ \footnotemark[1] & 2.820~eV \\
Dielectric constant $\varepsilon$ \footnotemark[1] & 9.0 \\
\hline
Elastic stiffness constants\footnotemark[2] & \\
$C_{11}$ & 8.26$\times $10$^{10}$~N/cm$^{ - 2}$ \\
$C_{12}$ & 4.98$\times $10$^{10}$~N/cm$^{ - 2}$\\
Deformation potential \footnotemark[2] & \\
$a$ & -4.25~eV\\
$b$ & -1.2~eV\\
Elastic strain $ \in $ \footnotemark[3] & 0.26~{\%} \\
Band gap corrections  due to strain \footnotemark[4] &  \\
Heavy-hole band $\Delta E_{hh} $ & 2~meV \\
Light-hole band $\Delta E_{lh} $ & 16~meV \\
\hline
Electron effective mass $m_e$ \footnotemark[5] & 0.15$m_0$ \\
Heavy-hole effective mass & \\
along growth direction $m_{hh}^z$ \footnotemark[6] & 0.8$m_0$ \\
\end{tabular}
\end{ruledtabular}
\footnotetext[1]{Ref.~\onlinecite{ref1}.}
\footnotetext[2]{Ref.~\onlinecite{ref2}.} \footnotetext[3]{$\in =
{\left[ {a_0 \left( {ZnSe} \right) - a_0 \left( {GaAs} \right)}
\right]} \mathord{\left/ {\vphantom {{\left[ {a_0 \left( {ZnSe}
\right) - a_0 \left( {GaAs} \right)} \right]} {a_0 \left( {GaAs}
\right)}}} \right. \kern-\nulldelimiterspace} {a_0 \left( {GaAs}
\right)}$.} \footnotetext[4]{Calculation according
Ref.~\onlinecite{ref3}.} \footnotetext[5]{Ref.~\onlinecite{ref4}.}
\footnotetext[6]{Because of uncertainty in heavy-hole effective
mass given in the literature (values are in the range 0.6$-$1.0),
for determination heavy-hole effective mass we take Luttinger
parameters $\gamma _1 = $2.45 $\gamma _2 = $0.61, which were
determined in Ref.~\onlinecite{ref5} from two-photon
magnetoabsorption measurements.}
\end{table}

{\it Parameters of barrier alloys}. Exciton energy of the barrier
materials ($E_X^b )$ has been evaluated directly from the exciton
resonance in reflectivity spectrum measured at $T$=1.6~K. Taking
for the exciton binding energy 20~meV (which is the value known
for ZnSe) we estimate the band gap of the barrier $E_g^b $, this
value is given in Table~\ref{tab1}. The band gap discontinuity to
ZnSe ($\Delta E_g )$ is also included in the table. In the model
calculations performed in Sec.~\ref{sec3} we will use values from
Table~\ref{tab1}, that are received from experiment.

To assign a certain composition of components in ternary and
quaternary barrier materials the following considerations have
been used. All structures were grown very closely lattice-matched
to GaAs substrates, as confirmed by X-ray measurements. This
condition gives us a ratio for the composition of different
components in the quaternary alloys. For Zn$_{1 -
x}$Mg$_{x}$S$_{y}$Se$_{1 - y}$ lattice-matched alloys the results
of Refs.~\onlinecite{ref6, ref7} have been used, which allow us to
assign the barrier with $\Delta E_g $=200~meV to
Zn$_{0.89}$Mg$_{0.11}$S$_{0.18}$Se$_{0.82}$ and the one with
$\Delta E_g $=100~meV to Zn$_{0.95}$Mg$_{0.05}$S$_{0.09}$Se$_{0.91
}$.

We give here more details for parameters we use for Zn$_{1 - x -
y}$Be$_{x}$Mg$_{y}$Se alloy parameterization, as literature data
are rather limited and give large scattering. The band gap
variation in ternary alloy Zn$_{1 - y}$Mg$_{y}$Se taken from
Ref.~\onlinecite{ref8},

\begin{eqnarray}
E_{g}(Zn_{1 - y}Mg_{y}Se)=
\quad \quad \quad \quad \quad \quad \quad \quad
\quad \quad \quad \quad \quad \nonumber\\
E_{g}(ZnSe) + 1.37y + 0.47y(y-1)
\label{eq2-1}
\end{eqnarray}
agrees well with results given by different groups. For the band
gap variation of Zn$_{1 - x}$Be$_{x}$Se alloys we utilize the
results of Ref.~\onlinecite{ref9}, where the full range of
contents from ZnSe to BeSe has been studied. The band gap of BeSe
at room temperature was determined as 5.55~eV and the respective
band gap variation for the alloy has been fitted by the following
equation

\begin{eqnarray}
E_{g}(Zn_{1 - x}Be_{x}Se)=
\quad \quad \quad \quad \quad \quad
\quad \quad \quad \quad \quad \quad \quad \nonumber\\
E_{g}(ZnSe) + 2.87x + 1.1x(x-1) .
\label{eq2-2}
\end{eqnarray}

For the relatively small values of cation substitution ($x$, $y$
$<$ 0.2) it is reasonable to construct the band gap variation of
the quaternary alloy Zn$_{1 - x - y}$Be$_{x}$Mg$_{y}$Se as a
linear combination of Eqs.~(\ref{eq2-1}) and (\ref{eq2-2})

\begin{eqnarray}
E_{g}(Zn_{1 - x - y}Be_{x}Mg_{y}Se)= E_{g}(ZnSe) + 2.87x
\quad \quad  \nonumber\\
+ 1.1x(x-1) + 1.37y + 0.47y(y-1) .
\label{eq2-3}
\end{eqnarray}

Lattice matching of the quaternary alloy to the lattice constant
of GaAs $a_{0}$(GaAs)=5.653~{\AA} at $T$=300~K gives us a
relationship for $x$ and $y$ ingredients of the alloy. Based on
the lattice constants for the binary alloys (5.6676~{\AA} for
ZnSe, 5.1520~{\AA} for BeSe and 5.904 {\AA} for MgSe \cite{ref1})
and the Vegard law \cite{ref57} the following dependencies for the
lattice constants of ternary alloys (giving in \AA) can be derived
(for $T$=300~K)

\begin{equation}
a_{0}(Zn_{1 - x}Be_{x}Se)= 5.6676 - 0.516x
\label{eq2-4}
\end{equation}

\begin{equation}
a_{0}(Zn_{1 - y}Mg_{y}Se)= 5.6676 + 0.236y .
\label{eq2-5}
\end{equation}

Lattice matching to GaAs corresponds to the ternary alloy
Zn$_{0.971}$Be$_{0.029}$Se and quaternary alloys satisfying a
condition

\begin{equation}
y = 2.186x - 0.062 .
\label{eq2-6}
\end{equation}

Combining Eqs.~(\ref{eq2-3}) and (\ref{eq2-6}) one can arrive at
the following condition for the energy gap of the lattice-matched
quaternary alloy

\begin{eqnarray}
E_{g}(Zn_{1 - x - y}Be_{x}Mg_{y}Se)=
\quad \quad \quad \quad \quad \quad \quad \quad
\quad \quad  \nonumber\\
E_{g}(ZnSe) + 3.346x^{2} + 3.61x - 0.054 . \label{eq2-7}
\end{eqnarray}

We derive the Be and Mg content in Zn$_{1 - x -
y}$Be$_{x}$Mg$_{y}$Se from experimental values of the enegy gap
and with use of Eqs.~(\ref{eq2-7}) and (\ref{eq2-6}). Respective
data are given in Table~\ref{tab1}.

{\it Experimental methods.} Photoluminescence (PL), PL excitation,
reflectivity (R) and spin-flip Raman scattering (SFRS)
spectroscopies were exploited for experimental study of trion
parameters. Optical spectra were detected at a low temperature
$T$=1.6~K. Different {\it cw} lasers were used for
photoexcitation, e.g. UV lines of an Ar-ion laser, a He-Cd laser
and a dye laser (Stylben~3). A halogen lamp was used in
reflectivity experiments. External magnetic fields were applied
along the structure growth axis (Faraday geometry). {\it dc}
magnetic fields up to 7.5~T were generated by a superconducting
solenoid and pulsed magnetic fields up to 47~T were used. In case
of {\it dc} field experiments direct optical access to the sample
was available through windows. For pulsed field experiments fiber
optics were used. In both cases circular polarization degree of
emitted/reflected light was analyzed. A complete set of
field-dependent PL spectra was collected during each magnetic
field pulse (for details see Ref.~\onlinecite{ref58}). Experiments
in a capacitor-driven 50~T mid-pulse magnet ($\sim $400~ms decay)
were performed at the National High Magnetic Field Laboratory (Los
Alamos, USA).


\section{\label{sec3}PROPERTIES OF CONFINED EXCITONS}

In the studies of trions, similar to excitons, the Rydberg energy
of the exciton in bulk semiconductor is often chosen as a
characteristic energy to parameterize the problem. In case of
quantum confined heterostructures it is also instructive to
compare the binding energies of trion states with the binding
energies of confined excitons. We will follow this tradition in
our investigations of charged excitons in ZnSe-based QW's.
Published information on the properties of excitons (e.g.
effective mass, binding energy, g-factor, radiative and
nonradiative dampings, {\it etc.}) in ZnSe-based QW's with binary
well material is rather limited. Therefore, we forestall the
results on trions in this section where the exciton parameters for
the ZnSe-based QW's will be evaluated from optical and
magneto-optical experiments and from variational calculations.

\subsection{\label{sub3A} Theoretical model for magneto-excitons}

In this section we will briefly describe the calculations of
exciton levels in the quantum wells in presence of magnetic field
directed along the growth axis.

All the calculations of exciton states presented in this paper
where made within a parabolic approximation, i.e. the admixture of
the light-hole states and all effects of nonparabolicity are
neglected. The quantization of the electron and hole states along
the structure growth axis ($z$-axis) provides us with the natural
basis in growth direction. We expand the wave function of the
exciton in a series

\begin{equation}
\psi (\rho,z_e,z_h ) = \sum {A_{i,j,n} \xi _i (z_e ) \zeta _j
(z_h) \psi_n (\rho )} ,
\label{eq3-1}
\end{equation}

$\xi _i (z_e )$, $\zeta _j(z_h )$ are the sets of solutions of
one-dimensional (1D) Schr\"{o}dinger equation for electron and
hole in z-direction. The choice of the radial basis $\psi _n (\rho
)$ will be discussed bellow.

The Eq.~(\ref{eq3-1}) represents the basis set for calculation the
exciton binding energy by diagonalization of the respective
matrix. In case of strong confinement (i.e. when the Coulomb
interaction is significantly less than the separation between
quantum confined states) one can neglect the excited single
particle states and the exciton problem reduces to 1D radial
equation with the Coulomb potential weighted over the ground
states of the electron and hole [for $i,j = $1 in
Eq.~(\ref{eq3-1})].

However, in the case of shallow (or wide) QW's, where energy
separation between levels of quantum confinement is small, such a
reduction of the basis in growth direction is not possible and to
calculate the spectrum of magneto-excitons a numerical
diagonalization scheme was used \cite{ref43}. Though the solutions
of the 1D radial equation are far from the real exciton wave
functions they form an orthonormalized basis that can also be used
in Eq.~(\ref{eq3-1}). Such choice of the radial expansion basis
$\psi _n (\rho )$ allows evaluation of exciton parameters for wide
range of magnetic fields including zero field limit, and provides
better results for shallow QW's than the simple one-dimensional
calculations. We present here the main line of this approach.

In the parabolic approximation, the QW electron-hole ($e-h$)
Hamiltonian in the magnetic field ${\bf B}$=(0,0,$B$) takes the
form

\begin{equation}
H = H_{ez} + H_{hz} + H_{2D} + U_{eh} \equiv H_0 + U_{eh}  .
\label{eq3-2}
\end{equation}

Here

\begin{equation}
H_{jz} = - \frac{\hbar ^2}{2m_j^z }\frac{\partial ^2}{\partial
z_j^2 } + V_j (z_j ),\,\,\,\,\,\,\,\,\,\,\,\,j = e,h .
\label{eq3-3}
\end{equation}

$V_j = \Delta E_C (\Delta E_V )$ is the band-offset potentials in
conduction (valence) band, $m_e^z$ and $m_h^z$ are effective
masses along growth direction of the electron and the heavy-hole,
respectively. We will not account for weak anisotropy of electron
effective mass in QW structures, i.e. take $m_e^z =m_e^{xy} =m_e$.

Hamiltonian $H_{2D}$ describes the two-dimensional (2D) motion of
a free electron-hole pair in the magnetic field,

\begin{eqnarray}
H_{2D} = \frac{1}{2m_e}\left( { - i\hbar \nabla {\rm {\bf \rho
}}_e +\frac{e}{c}{\rm {\bf A}}_e } \right)^2
\quad \quad \quad \quad \quad \quad  \quad \nonumber\\
+ \frac{1}{2m_{h}^{xy} }\left( { - i\hbar \nabla {\rm {\bf \rho
}}_h - \frac{e}{c}{\rm {\bf A}}_h } \right)^2 , \label{eq3-4}
\end{eqnarray}

where ${\rm {\bf A}}_j = \textstyle{1 \over 2}{\rm {\bf B}}\times
{\rm {\bf \rho }}_j $ is the vector potential in the symmetric
gauge, ${\rm {\bf \rho }}_{e(h)} $ are the in-plane coordinates of
electron (hole), ${\rm {\bf \rho }} = {\rm {\bf \rho }}_e - {\rm
{\bf \rho }}_h = (x,y)$. We neglect here the mass difference in
the well and barrier layers. The potential

\begin{equation}
U_{eh} (\rho ,\,z_e ,\,z_h ) = - \frac{e^2}{\varepsilon }\left[
{\frac{1}{\sqrt {\rho ^2 + \left( {z_e - z_h } \right)^2} }}
\right]
\label{eq3-5}
\end{equation}

is the Coulomb interaction between the electron and hole.
$\varepsilon $ is the dielectric constant.

The matrix elements of the Hamiltonian (\ref{eq3-2}) in the basis
(\ref{eq3-1}) can be written in the following form

\begin{eqnarray}
H_{i'j'n'}^{ijn} = (E_i^e + E_j^h + E_n^X )\delta _{ii'} \delta
_{jj'} \delta _{nn'}
\quad \quad \quad \quad \quad   \nonumber\\
+ < ijn\vert U_{eh} \vert i'j'n' > - \delta _{ii'} \delta _{jj'} <
n\vert U_{11}^{11} \vert n' > .
\label{eq3-6}
\end{eqnarray}

Here $E_i^e ,E_j^h $ are quantum confined energies of electrons
and holes, $E_n^X $ are the eigenvalues of the radial exciton
equation with Coulomb potential averaged over the ground electron
and hole states $U_{11}^{11} (\rho )$. Here we have made use of
the fact that basis functions $\psi _n (\rho )$ are the
eigenfunctions of radial exciton Hamiltonian for the {\bf K}=0
$s$-exciton (with the angular momentum projection of the relative
$e-h$ motion $l_z = 0)$ that reads as

\begin{eqnarray}
H_\rho \psi _n (\rho )\, = \,E_n^X \psi _n (\rho ),
\nonumber\\
H_\rho = - \frac{\hbar ^2}{2\mu }\nabla _\rho ^2 +
\frac{e^2B^2}{8\mu c^2}\rho ^2 + U_{11}^{11} (\rho ),
\label{eq3-7}
\end{eqnarray}

where $\mu = \left( {1 \mathord{\left/ {\vphantom {1 {m_e }}}
\right. \kern-\nulldelimiterspace} {m_e } + 1 \mathord{\left/
{\vphantom {1 {m_{h}^{xy} }}} \right. \kern-\nulldelimiterspace}
{m_{h}^{xy} }} \right)^{ - 1}$ is the reduced exciton mass, and

\begin{eqnarray}
U_{ij}^{i'j'} (\rho ) = \int {dz_e \int {dz_h}} \quad \quad \quad
\quad \quad \quad \quad
\quad \quad \quad \quad  \nonumber\\
U_{eh} (\rho ,\,z_e ,\,z_h )\, \xi _i (z_e
)\xi _{i'} (z_e )\,\xi _j \,(z_h )\xi _{j'} \,(z_h ),
\label{eq3-8}
\end{eqnarray}

\begin{equation}
< ijn\vert U_{eh} \vert i'j'n' >~ = \int d ^2\rho \;\;
U_{ij}^{i'j'} (\rho )\psi _n (\rho )\psi _{n'} (\rho ) .
\label{eq3-9}
\end{equation}

The diagonalization of Hamiltonian (\ref{eq3-6}) provides us with
both the eigenvalues and eigenfunctions of the exciton states. The
dimension of the basis used depends on the relative values of the
quantization energy and the Coulomb interaction. The wider is the
QW and, consequently, the smaller is the vertical quantization the
larger should be the number of $z$-functions taken into account.
In our calculation we use all quantum confinement states in real
QW and ten radial basis functions. The parameters required for the
calculations are the QW width ($L_z )$ and band offsets ($\Delta
E_C $, $\Delta E_V )$, effective masses of electron and hole in
vertical direction ($m_e $, $m_{h}^z )$, in-plane reduced exciton
mass ($\mu )$ and dielectric constants ($\varepsilon )$. We took
$\Delta E_C $, $\Delta E_V $, $m_e $, $m_{h}^{z}$ (namely
$m_{hh}^{z}$ for the heavy-hole exciton) and $\varepsilon$ from
the literature, $\mu $ and $L_z $ were obtained experimentally.
The primary parameters are given in Tables~\ref{tab1}, \ref{tab2}
and determined ones are summarized in Tables~\ref{tab3} and
\ref{tab4}.

In the frame of this approach we have calculated exciton energies
for the studied QW structures {\it vs} QW width, exciton binding
energies and modification of these parameters in external magnetic
fields up to 50~T. Results of these calculations are included in
Figs.~\ref{fig4} and \ref{fig5}.

\begin{figure}
\includegraphics{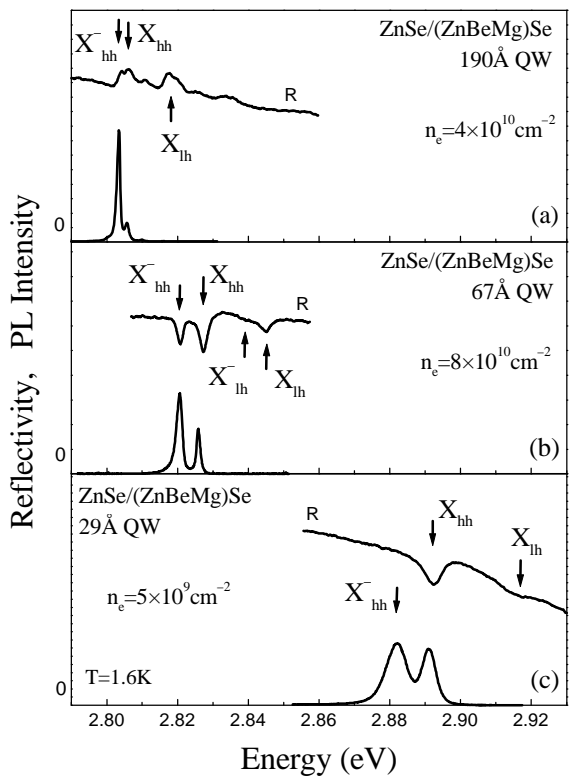}
\caption{\label{fig2} Reflectivity and photoluminescence spectra
taken from ZnSe/Zn$_{0.82}$Be$_{0.08}$Mg$_{0.10}$Se (type~A)
structures with different QW widths of 190~{\AA} (a), 67~{\AA} (b)
and 29~{\AA} (c). Arrows indicate the heavy- hole exciton
($X_{hh}$), the light-hole exciton ($X_{lh}$), and the negatively
charged excitons ($X_{hh}^{ - }$ and $X_{lh}^{ - }$). Electron
concentration $n_e$ in QW is given in the figure. $T$=1.6~K.}
\end{figure}

\subsection{\label{sub3B} Optical spectra of excitons}

Figure~\ref{fig2} displays typical optical spectra for three
ZnSe/Zn$_{0.82}$Be$_{0.08}$Mg$_{0.10}$Se single QW's, which covers
the whole range of the studied QW widths $L_z $ from 29 to
190~{\AA}. Photoluminescence and reflectivity spectra were
measured in the absence of external magnetic fields at a
temperature of 1.6~K. Exciton resonances corresponding to the
states formed with heavy- and light holes ($X_{hh}$ and $X_{lh}$,
respectively) are clearly visible in reflectivity spectra. Trion
resonances shifted to low energies from the $X_{hh}$ energy are
seen in 67 and 190~{\AA} QW's. Their intensities in reflectivity
spectra are proportional to the electron densities \cite{ref12}.
Low electron concentration and relatively large broadening make
the trion resonance unresolvable for a 29~{\AA} QW. For all
structures shown, PL spectra consist of two lines, where the
low-energy line is due to the radiative recombination of
negatively charged excitons ($X_{hh}^ - )$ and the high-energy
lines is due to recombination of neutral excitons ($X_{hh})$.
Details of their identification in ZnSe-based QW's can be found in
Ref.~\onlinecite{ref11}. With decreasing QW width, an increase of
confined energies of carriers in the conduction and valence bands
causes the high-energy shift of exciton transitions. It is
accompanied by the broadening of exciton transitions due to QW
width and barrier alloy fluctuations.

\begin{figure}
\includegraphics{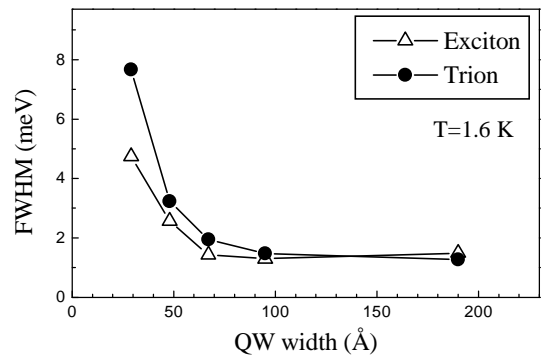}
\caption{\label{fig3} A full width at a half maximum (FWHM) of
trion (circles)- and exciton (triangles) PL lines as a function of
QW width in ZnSe/Zn$_{0.82}$Be$_{0.08}$Mg$_{0.10}$Se structures.
Lines are guides to the eye.}
\end{figure}

The full width at a half maximum (FWHM) of exciton and trion PL
lines is plotted in Fig.~\ref{fig3}. Exciton linewidth for
$L_z>$67~{\AA} is smaller than 1.2~meV, which evidence the high
structural and optical quality of the studied samples. For this
range of QW width the trion linewidth roughly coincides with the
exciton one. It is interesting that for $L_z<$100~{\AA}, exciton
PL lines are narrower than the trion lines. The difference
achieved 60{\%} in a 29~{\AA} QW. Possible reasons for that we
will discuss below in Sec.~\ref{sub5D}.

\subsection{\label{sub3C} Excitons in high magnetic fields}

Application of external magnetic fields allows evaluating
important exciton parameters such as the in-plane reduced
effective mass $\mu $, and the g factor that characterizes the
spin splitting of excitons due to the Zeeman effect. These
parameters are important for understanding and calculating the
spin- and energy structure of trions and excitons. Experiments
were performed in pulsed magnetic fields to 47~Tesla. Application
of high magnetic fields was required to induce sufficient energy
shift of strongly bound excitons in ZnSe-based QW's.
Photoluminescence spectra were measured in two circular
polarizations corresponding to two spin states of optically active
excitons. Evolution of PL spectra with increasing magnetic fields
is discussed in detail in Refs.~\onlinecite{ref16, ref10}. Results
on the spin splitting of excitons will be presented and discussed
in Sec.~\ref{sub3E}. Here we concentrate on the energy shift of
exciton with increasing magnetic fields.

\begin{figure}
\includegraphics{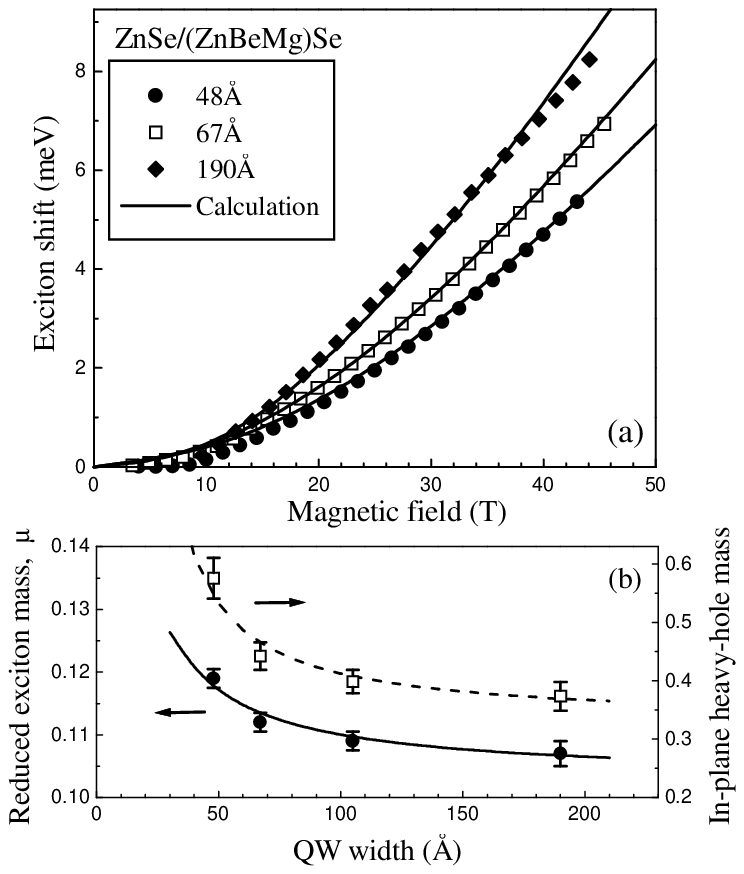}
\caption{\label{fig4} (a) Exciton energy {\it vs} magnetic field
strength for different type A QW's: 190~{\AA} (diamonds), 67~{\AA}
(squares) and 48~{\AA} (circles). Center-of-gravity of the exciton
spin doublet in PL spectra is taken for the exciton energy. $T =
$1.6~K. Lines show results of model calculations.\\
(b) Reduced mass of exciton $\mu$ (left axis) and in-plane
heavy-hole mass $m_{hh}^{xy}$ (right axis) {\it vs} QW width.
Symbols are experimental data. A solid line is an interpolation of
data points by the hyperbolic function $\mu = ( 0.103 + 0.7 /
L_z$[{\AA}]$)m_0$. A dashed line is a result of calculation along
$m_{hh}^{xy} = m_e / {\left( \mu m_e - 1  \right)}$. }
\end{figure}

To avoid spin splittings a center-of-gravity of the exciton spin
doublet was evaluated and plotted as a function of magnetic field
strength in Fig.~\ref{fig4}a for QW's of different widths.
Characteristic diamagnetic shift of excitons is seen for all
samples. Exciton shift increases in wider QWs, which coincides
with decreasing binding energy of excitons.

\begin{table*}
\caption{\label{tab3}Effective masses and parameters of Zeeman
splitting for excitons and free carriers in ZnSe-based QW's.
Values for effective masses are given in $m_0$. All experimental
data, except $g_e^{xy}$, were measured in the Faraday
configuration.}
\begin{ruledtabular}
\begin{tabular}{cc|ccc|cc|cc|c|cc|cc}
\multicolumn{2}{c}{} &\multicolumn{3}{c}{Structure parameters} &\multicolumn{2}{c}{Effective mass} &\multicolumn{7}{c}{g factor} \\
SAMPLE & Type & $L_z$ & $\Delta E_g$& $\Delta _{hh - lh}$& $\mu$ & $m_{hh}^{xy}$ &\multicolumn{2}{c}{$g_X$} & \multicolumn{1}{c}{$g_e^{xy}$} &\multicolumn{2}{c}{$g_e^z$} &\multicolumn{2}{c}{$g_{hh}$\footnotemark[2]}   \\
& & ({\AA}) & (meV) & (meV) &  &  &  $<$10~T& 40~T & SFRS & SFRS & Calc.\footnotemark[1] & $<$10~T & 40~T \\
\hline
cb1175 & A & 29  & 240 & 24 &       &      &     &     & 1.21 & 1.17 & 1.13 &     &     \\
cb1173 & A & 48  & 240 & 20 & 0.119 & 0.58 & 1.0 & 1.3 & 1.17 & 1.17 & 1.13 & 2.2 & 2.5 \\
cb1041 & A & 67  & 240 & 18 & 0.112 & 0.44 & 0.4 & 0.9 &      &      & 1.13 & 1.5 & 2.0 \\
cb1174 & A & 67  & 240 & 17 &       &      &     &     & 1.13 & 1.13 & 1.13 &     &     \\
cb1198 & A & 95  & 240 & 13 &       &      &     &     & 1.11 & 1.11 & 1.12 &     &     \\
cb1172 & A & 190 & 240 & 11 & 0.107 & 0.37 & 0.4 & 0.4 & 1.11 & 1.11 & 1.12 & 1.5 & 1.5 \\
\hline
zq1038 & B & 80  & 200 & 17 &       &      &     &     & 1.17 & 1.14 & 1.15 &     &     \\
zq1113 & B & 105 & 200 & 16 & 0.109 & 0.40 & 0.5 & 0.7 &      &      & 1.14 & 1.6 & 1.8 \\
\hline
zq703  & C & 45  & 100 & 19 & 0.115 & 0.50 & 0.4 & 0.8 &      &      & 1.22 & 1.6 & 2.0 \\
\end{tabular}
\end{ruledtabular}
\footnotetext[1]{From our calculation.} \footnotetext[2]{Evaluated
as $g_{hh}=g_e+g_X$, where $g_e$ is corresponding $g_e^z$ measured
by SFRS or calculated in case if no experimental value is
available.}
\end{table*}

Solid lines in Fig.~\ref{fig4}a show the best fit of experimental
data in the frame of the model described in Sec.~\ref{sub3A}.
Parameters used for the calculations were taken from
Tables~\ref{tab1} and \ref{tab2}. The exciton reduced mass $\mu $
is the only free parameter in the fit. Determined values of $\mu $
are included in Table~\ref{tab3} and also plotted in
Fig.~\ref{fig4}b {\it vs} $L_z $. The reduced mass increases for
thinner QW's with functional dependence that can be interpolated
as $\mu = ( 0.103 + 0.7 / L_z$[{\AA}]$)m_0$ (see solid line in
Fig.~\ref{fig4}b). Taking the value of the electron effective mass
$m_e = 0.15m_0$ to be independent of the well width (this is valid
as confinement energies in the studied structures are small, do
not exceed 60~meV, and one should not expect strong contribution
of nonparabolicity in the conduction band to the electron
effective mass), the in-plane values for the heavy-hole effective
mass $m_{hh}^{xy} $ are determined by means of relationship
$m_{hh}^{xy} \, = \,\mu {m_e } \mathord{\left/ {\vphantom {{m_e }
{\left( {m_e \, - \,\mu } \right)}}} \right.
\kern-\nulldelimiterspace} {\left( {m_e \, - \,\mu } \right)}$.
These values are displayed in Fig.~\ref{fig4}b by open squares
(right axis). One can see that $m_{hh}^{xy} $ increases
significantly from 0.37~$m_{0}$ in a 190~{\AA} QW to 0.58~$m_{0}$
in a 48~{\AA} QW. This fact should be accounted in model
calculations of trions binding energies {\it vs} QW width (e.g.
\cite{ref36}).

\subsection{\label{sub3D} Confined excitons}

\begin{figure}
\includegraphics{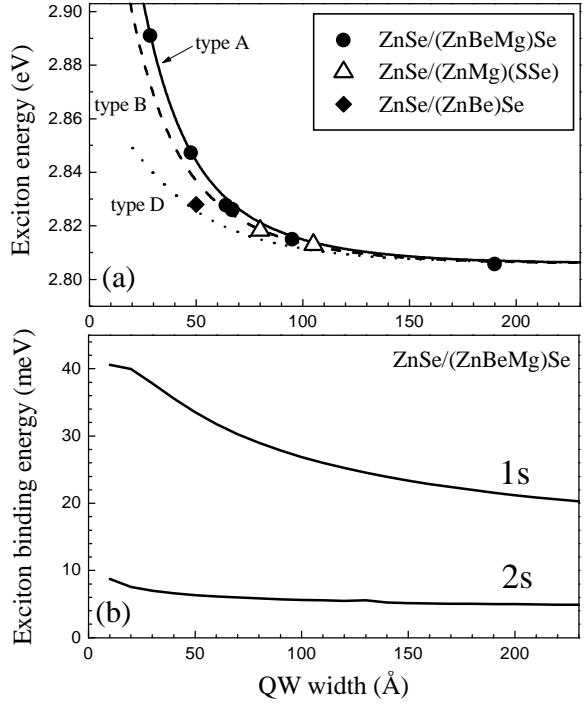}
\caption{\label{fig5} (a) Lines are exciton energies calculated
for different types of structures as a function of QW width.
Symbols present experimental data. The nominal (i.e.
technological) values of QW width were slightly corrected to put
experimental points at calculated dependencies. \\
(b) The calculated exciton binding energy (1s- and 2s states) as a
function of QW width for type~A
ZnSe/Zn$_{0.82}$Be$_{0.08}$Mg$_{0.10}$Se structures. }
\end{figure}

QW width dependencies for exciton energy $E_X $ and exciton
binding energy $E_B^X $ were calculated by means of the model
described in Sec.\ref{sub3A}. Structure parameters used for the
calculations are in Tables~\ref{tab1}, \ref{tab2} and \ref{tab3}.
Results of these calculations are displayed in Fig.~\ref{fig5}.

In Fig.~\ref{fig5}a calculated exciton energies for different
types of structures (which differ by barrier heights and band
offsets) are plotted by lines. We use these dependencies and
experimental values of the exciton energies to determine QW width
in the studied structures. Experimental data are shown by symbols
and included in Table~\ref{tab4} . Nominal values of the QW width
evaluated from the technological parameters coincide with high
accuracy with the calculated values.

In Fig.~\ref{fig5}b binding energies of 1s and 2s exciton states
are plotted as a function of $L_z $ for
ZnSe/Zn$_{0.82}$Be$_{0.08}$Mg$_{0.10}$Se structures. One can see
that in ZnSe QW's, the exciton binding energy $E_B^X $ (i.e. the
binding energy of 1s state) has its maximum for QW's with $L_z
\approx $20~{\AA}. This value is 40~meV, i.e. about twice as large
as the bulk exciton Rydberg $R$=20~meV. This value indicates that
in the ZnSe QW's, the exciton is quasi-two-dimensional, as its
binding energy is considerably smaller than the binding energy in
2D limit, 4$R$=80~meV.

\subsection{\label{sub3E} Zeeman splitting of excitons and free carriers}

The spin splitting of the exciton states is composed of the
splitting of conduction- and valence bands which are characterized
by the electron- ($g_e )$ and hole ($g_{hh} $ or $g_{lh} $ for the
heavy-hole or light-hole subbands) gyromagnetic ratios (g values).
Following Ref.~\onlinecite{ref44} we define the exciton g factor
as $g_X = g_{hh} - g_e $. Values of g factors depend on the band
structure parameters. They can be calculated with high accuracy
for the conduction band. Modeling of the g factors for the valence
band is more complicated due to the mixing of heavy-hole and
light-hole bands, whose splittings depend on the structure
parameters (strain, quantum confinement, differences in band
parameters of the barrier and QW materials) (see e.g.
Ref.~\onlinecite{ref53}). Detailed investigations of electron g
factors in (Zn,Mg)Se and Zn(S,Se) alloys \cite{ref52} and
ZnSe/(Zn,Mg)(S,Se) QW structures [51] have been performed by means
of spin-flip Raman scattering spectroscopy. It was shown that with
a properly chosen set of band parameters, the five-band model
calculations, which accounts for the off-diagonal spin-orbit
coupling terms (TO model in Ref.~\onlinecite{ref52}), give a very
good agreement for the band gap dependence of $g_e $ in ZnSe-based
ternary alloys. The $g_e $ values for the quaternary alloy
(Zn,Mg)(S,Se) are also in a reasonable agreement with model
estimations \cite{ref52, ref51}. Estimation for the barrier
materials in our structures gives us $g_e $=+1.32 and +1.38 for
Zn$_{0.95}$Mg$_{0.05}$S$_{0.09}$Se$_{0.91}$ and
Zn$_{0.89}$Mg$_{0.11}$S$_{0.18}$Se$_{0.82 }$, respectively. These
values exceed the electron g factor in ZnSe $g_e $=+1.12 (see
Ref.~\onlinecite{ref52} and references therein).

\begin{figure}
\includegraphics{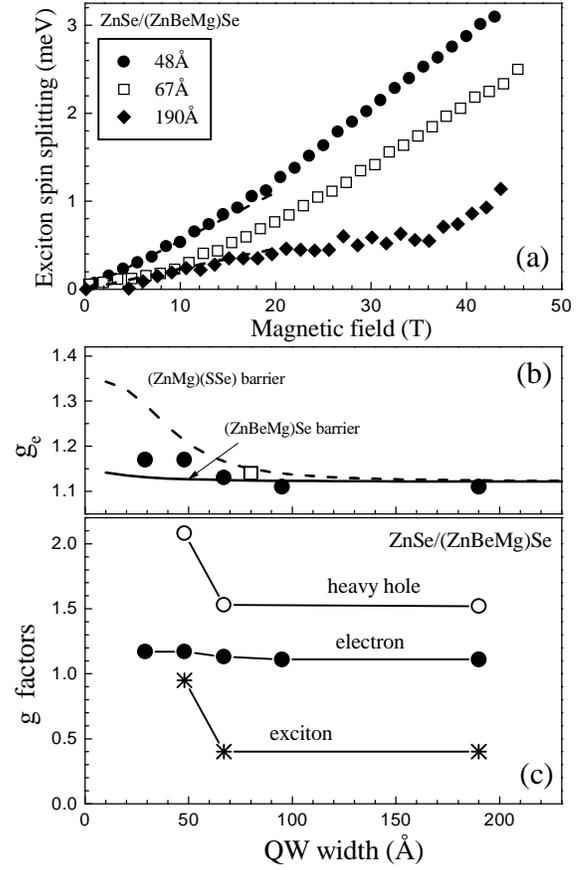}
\caption{\label{fig6} (a) Zeeman splitting of exciton as a
function of magnetic field for different QW widths: 190~{\AA}
(diamonds), 67~{\AA} (squares) and 48~{\AA} (circles). \\
(b) Electron g factors $g_e^z $ for QW's with type A (circles and
solid line) and type B (squares and dashed line) barriers: lines
are calculation and symbols are experimental data measured by
spin-flip Raman scattering in the Faraday geometry. \\
(c) g factors in ZnSe/Zn$_{0.82}$Be$_{0.08}$Mg$_{0.10}$Se QW's.
Exciton values for the low field limit ($B<$10T) are shown by
stars. Solid circles present the electron g factor $g_e^z $.
Heavy-hole g factor evaluated as $g_{hh} = g_e + g_X $ is given by
open circles. Lines are guides to the eye.}
\end{figure}

A well width dependence of the electron g factor in
ZnSe/Zn$_{0.89}$Mg$_{0.11}$S$_{0.20}$Se$_{0.80}$ QW's (which are
practically identical to our samples with
Zn$_{0.89}$Mg$_{0.11}$S$_{0.18}$Se$_{0.82}$ barriers and $\Delta
E_g \approx $200 meV) has been investigated in
Ref.~\onlinecite{ref51}. Comparison with the model calculations
allows authors to conclude that in contrast to the alloys the
three-band model is sufficient to calculate g factors in QW
structures. Two additional factors should be accounted for in
QW's. First, there is difference in the band parameters in the
well and barrier materials. In ZnSe-base QW's with relatively low
barriers with band gaps in the region of 2.8-3.1~eV, the main
contribution comes from the difference in the spin-orbit splitting
$\Delta _0 $. However, the change of $\Delta _0 $ is rather small
in the case of the group II element to be substituted, but is
large if the group VI element is altered. In other words, one can
omit the variation of $\Delta _0 $ for the structures with
(Zn,Be,Mg)Te barriers, but it should be accounted for in the case
of (Zn,Mg)(S,Se) barriers. The second consideration specific to
the QW structures is due to the anisotropy of the electron g
factors. Its value is relatively small (usually less than 10{\%}
of $g_e )$ and is controlled by the splitting of light-hole and
heavy-hole states due to strain and confinement effects.
Experimental values of $g_e^z $ and $g_e^{xy} $ obtained by
spin-flip Raman scattering for the type A and B ZnSe QW's confirm
the small value for $g_e $ anisotropy (see Table~\ref{tab3}).

Electron g factor for the type A QW's was measured by SFRS at the
University of Bath. Details of experimental technique are
published in Refs.~\onlinecite{ref51,ref52}. Experiments were
performed in the Faraday and Voigt geometries to measure $g_e^z $
and $g_e^{xy} $ components of the electron g factor, respectively.
Their values are included in Table~\ref{tab3} and $g_e^z $ are
shown by solid circles in Figs.~\ref{fig6}b and \ref{fig6}c.
Experimental data for an 80~{\AA}
ZnSe/Zn$_{0.89}$Mg$_{0.11}$S$_{0.18}$Se$_{0.82}$ QW were taken
from Ref.~\onlinecite{ref13} (shown by squire in
Fig.~\ref{fig6}b).

Using the results of Ref.~\onlinecite{ref51} we calculated average
values for $g_e $ in the structures of type A and type B. For the
type A structures with (Zn,Be,Mg)Se barriers we use equation~(7)
from Ref.~\onlinecite{ref51}

\begin{equation}
g_e (L_z )\,\, = \,\,g_e^{ZnSe} \, + \,\frac{4\gamma \,E_P
\,\Delta _0 }{3E_X^3 } . \label{eq3-10}
\end{equation}

Here $\,\,g_e^{ZnSe} \,$=+1.12 is the electron g factor in ZnSe,
$\gamma = \,E_X + E_B^X \, - \,E_g (ZnSe)\,$, $E_P =
23.4\,\,\mbox{eV}$ is the squared momentum matrix element in ZnSe,
$\Delta _0 \, = \,0.42$ eV in ZnSe, $E_X $ is exciton energy
measured experimentally (see Table~\ref{tab4}). The results are
shown in Fig.~\ref{fig6}b by a solid line. One can see that the
$g_e $ value is rather weakly dependent on the QW width for the
whole studied range from 29 to 190~{\AA}.

A more elaborate approach, accounting for the difference in
$\Delta _0 $ for QW and barriers layers, was applied for the
calculation of electron g factors in the type B QW's with
(Zn,Mg)(S,Se) barriers. It is described by Eq.~(8) from
Ref.~\onlinecite{ref51} and we do not detailed it here. In this
case $g_e $ is more sensitive to the QW width for $L_z <$100~{\AA}
(see dashed line in Fig.~\ref{fig6}b).

Hole g factors are strongly anisotropic (e.g. in-plane component
of $g_{hh} \approx 0)$ and their values are determined in a
complicated manner on the splitting of heavy-hole and light-hole
states $\Delta _{hh - lh}$ (see e.g. Ref.~\onlinecite{ref53}). We
are not aware of any simple calculation approach to this problem
and will limit ourselves by experimental dependencies. Note that
the bulk relation $g_{hh} = 3g_{lh} $ is not valid anymore in
QW's.

Exciton Zeeman splittings in QW's of different thickness in
Fig.~\ref{fig6}a show reasonably good linear dependence on
magnetic fields at $B<$10 T and some deviation from a linear
behavior at high fields. It is well known that the nonlinearity of
the exciton g factor is caused by the nonlinearity of its hole
component $g_{hh} $, that in turn is due to the admixing of
light-hole states in high magnetic fields. Electron Zeeman
splittings, as a rule, shows a linear dependence over a wide range
of magnetic fields, and this is true for ZnSe-based QW's
\cite{ref51}. To quantify the nonlinear spin splitting values we
include in Table~\ref{tab3} two values for the exciton ($g_X )$
and heavy-hole g factors estimated from linear interpolation of
data points at low magnetic fields ($B<$10~T) and evaluated from
the exciton spin splitting at $B$=40~T.

Exciton and carrier g factors for different QW's are collected in
Table~\ref{tab3} and displayed in Fig.~\ref{fig6}c. In the figure
the data are given for ZnSe/Zn$_{0.82}$Be$_{0.08}$Mg$_{0.10}$Se
QW's. Exciton values for the low field limit ($B<$10~T) are shown
by stars. In 190 and 67~{\AA} QW's they are equal to +0.4 and $g_X
$ increases to +1.0 in a 48~{\AA} QW. Solid circles trace the
electron g factor $g_e^z $, which is weakly dependent on the well
width. Heavy-hole g factor evaluated as $g_{hh} = g_e + g_X $ is
given by open circles. Its dependence on the QW width reflects the
$g_X $ behavior.

Now we have all parameters necessary for analysis of the spin- and
energy structure of the trion states reported in this paper. Some
further information on the exciton properties including radiative-
and nonradiative damping, coherent- and recombination dynamic of
excitons and trions in ZnSe-based QW's can be found in
Refs.~\onlinecite{ref12,ref33,ref14}. We turn now to the main part
of the paper, where the properties of charged excitons are
investigated.


\begin{figure}
\includegraphics{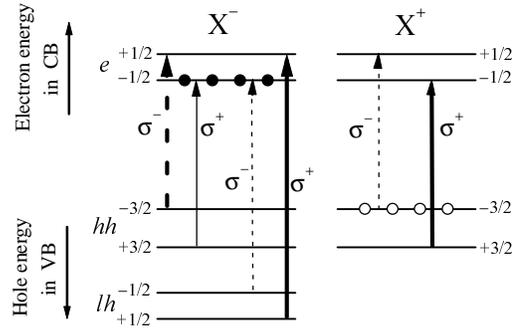}
\caption{\label{fig7} Scheme of the optical transitions in
ZnSe-based QW's for the creation of charged excitons in the cases
of completely polarized 2DEG (solid circles) and 2DHG (open
circles) induced by external magnetic fields. Optically-active
circular-polarized transitions are shown by arrows. The thick
arrows represent the transitions in which the trion formation is
allowed.}
\end{figure}

\section{\label{sec4}CHARGED EXCITONS}

\subsection{\label{sub4A} Identification of negatively- and positively charged
excitons}

Charged exciton states in optical spectra can be identified by
their specific polarization properties in external magnetic
fields. Analysis of the circular polarization degree of
photoluminescence is rather complicated. In addition to the spin
polarization of the free carriers, the spin-dependent trion
formation and spin relaxation of trions are involved \cite{ref13}.
However, polarization properties of the trion states in
reflectivity, absorption or transmission spectra allow to
distinguish trions from excitons and positively- and negatively
charged excitons from each other. Here we present in short
principles of the identification, and further details can be found
in Refs.~\onlinecite{ref11,ref12}.

\begin{figure}
\includegraphics{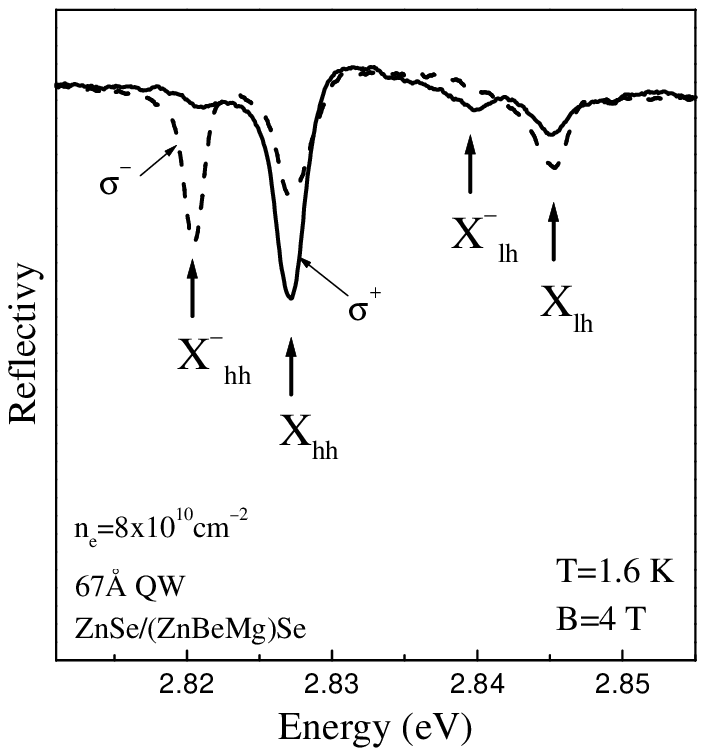}
\caption{\label{fig8} Reflectivity spectra taken from a 67~{\AA}
ZnSe/Zn$_{0.82}$Be$_{0.08}$Mg$_{0.10}$Se QW detected in different
circular polarization at a magnetic field of 4~T. Strongly
polarized resonances of negatively charged excitons related to the
heavy-hole ($X_{hh}^{ - }$) and the light-hole ($X_{lh}^{ - }$)
excitons are labeled by arrows. Electron concentration in QW is
$n_e$=8$\times $10$^{10}$~cm$^{ - 2}$, $T$=1.6~K.}
\end{figure}

The polarization degree of the trion resonance in reflectivity
spectra mirrors the polarization of free carriers in QW's which is
caused by thermal distribution of the carriers among the Zeeman
sublevels. This is due to the singlet spin structure of the trion
ground state, i.e. spins of two carriers with the same charges
(electrons in $X^{ - } $ and holes in $X^{ + })$ in the trion
complex are oriented antiparallel. A triplet trion state with
parallel orientation of these spins is unbound at zero magnetic
field and becomes bound in high fields only \cite{ref16}. When
free carriers are fully polarized by magnetic field the trions can
be excited optically only for one circular polarization. In case
of negatively charged excitons in ZnSe QW's with a positive
electron g factor it is $\sigma ^ - $ polarization.

A scheme of the optical transitions responsible for trion
excitation in strong magnetic fields is presented in
Fig.~\ref{fig7}. Spin-split states at the bottom of the conduction
band and the top of the valence band are shown. Arrows indicate
optical transitions, where the absorbed light promotes an electron
from the valence band to the conduction band and forms an exciton.
Exciton generation in the presence of free carriers results in
trion formation. Thick and thin arrows mark the allowed and
forbidden transitions for the trion excitation in its ground
state, when the carriers are fully polarized. It is clear from the
scheme that $X^{ - }$ related to the heavy-hole and light-hole
excitons will appear in opposite polarizations. Also $X^{ - }$ and
$X^{ + }$ in ZnSe QW's, where electron and hole g factors are
positive, can be clearly distinguished by their opposite
polarizations.

\begin{figure}
\includegraphics{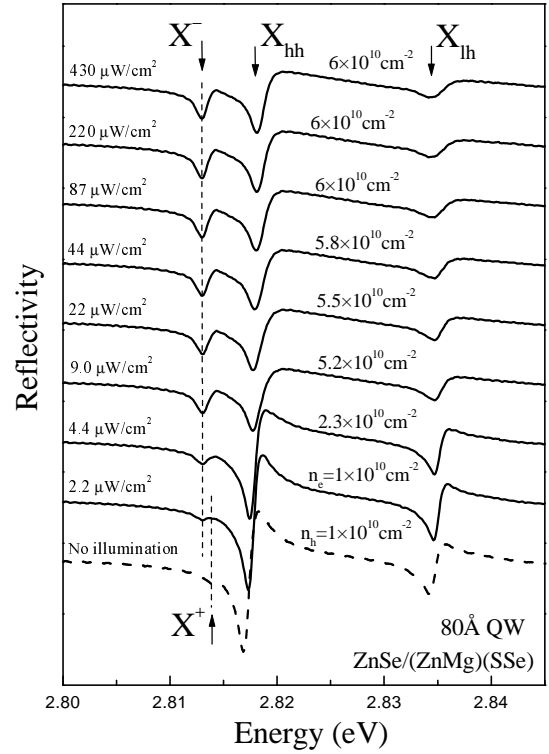}
\caption{\label{fig9} Tuning of electron (hole) gas concentration
by additional above-barrier illumination. The figure presents
reflectivity spectra of an 80~{\AA}
ZnSe/Zn$_{0.89}$Mg$_{0.11}$S$_{0.18}$Se$_{0.82}$ QW {\it vs}
illumination intensity of Ar-ion laser (3.5 eV). The laser power
is given in the figure. $B$=0~T, $T$=1.6~K.}
\end{figure}

In Fig.~\ref{fig8} typical reflectivity spectra containing
strongly polarized resonances of negatively charged excitons
associated with heavy-hole excitons ($X_{hh}^{ - })$ and
light-hole excitons ($X_{lh}^{ - })$ are given. Results are shown
for a 67~{\AA} ZnSe/Zn$_{0.82}$Be$_{0.08}$Mg$_{0.10}$Se QW and a
magnetic field of 4~T. In accordance with the selection rule
discussed above, $X_{hh}^{ - }$ and $X_{lh}^{ - }$ resonances show
up in different polarizations, i.e. $\sigma ^ - $ and $\sigma ^ +
$ respectively.

\begin{figure}
\includegraphics{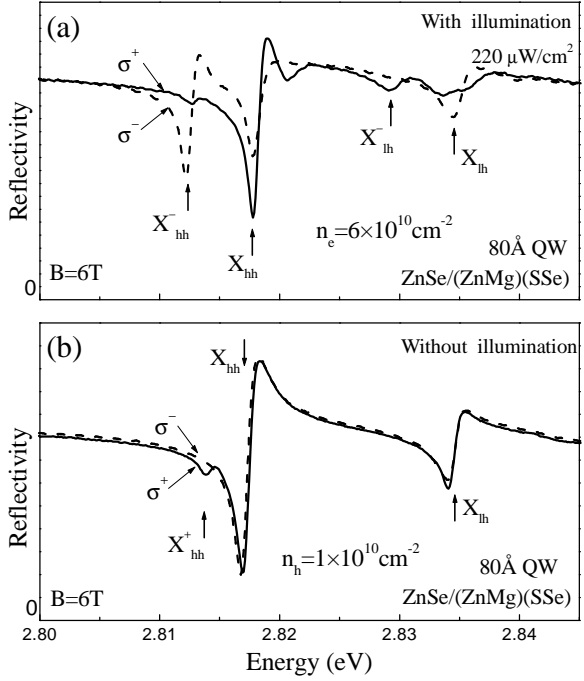}
\caption{\label{fig10} Reflectivity spectra of an 80~{\AA}
ZnSe/Zn$_{0.89}$Mg$_{0.11}$S$_{0.18}$Se$_{0.82}$ QW: (a) With
above-barrier illumination detected at different circular
polarization at a magnetic field of 6~T. Arrows indicate
negatively charged exciton ($X^{ - })$ and excitons formed
with heavy-hole and light-hole.\\
(b) Without above-barrier illumination at a magnetic field of 6~T.
Arrows indicate positively charged exciton ($X^{ + })$ and
excitons formed with heavy-hole and light-hole. $T$=1.6~K.}
\end{figure}

Examples of the opposite polarization of $X^{ - }$ and $X^{ + }$
in ZnSe QW's can be found in Ref.~\onlinecite{ref11} and in
Fig.~\ref{fig10} in the next section, where a recharging effect of
the QW by above-barrier illumination is discussed.

\subsection{\label{sub4B} Optical tuning of carrier density in QW's}

Optical tuning of a carrier density in QW's is a very reliable
method that is widely exploited for the trion studies
\cite{ref45,ref46,ref47,ref48}. Different structure designs have
been suggested for this purpose. The principle of the method is in
spatial separation of electron-hole pairs photogenerated by
photons with energies exceeding the barrier band gap. Depending on
the structure design, one type of carrier is captured by the
surface states, trapped centers in barriers or additional quantum
well. The other type of carrier is collected into the quantum well
where it is involved in the trion formation. Free carrier
concentration in the QW is tuned by the intensity of the
above-barrier illumination. However, the dependence of the
concentration on the illumination intensity can be very nonlinear
with a pronounced saturation at higher intensities. The optical
method can be also used for a fine tuning of carrier densities in
structures with modulation doping and/or under applied gated
voltage.

\begin{figure}
\includegraphics{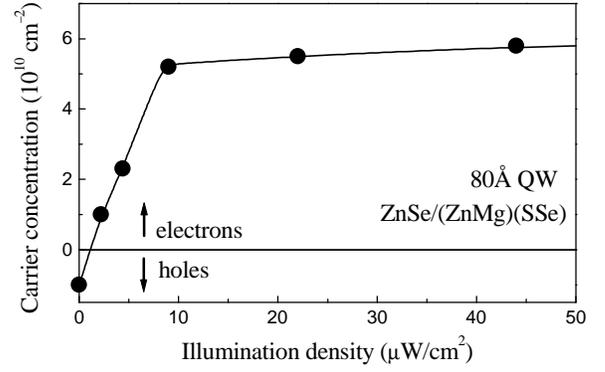}
\caption{\label{fig11} Electron (hole) concentration in the QW as
a function of illumination intensity. Symbols are experimental
data (for details see Fig.~9), solid line is a guide to the eye.}
\end{figure}

We will show here that optical tuning is very effective for the
types A and B of ZnSe-based heterostructures investigated in this
paper. Let us start with the type B structure zq1038, where the
80~{\AA} QW is separated from the surface by a 600~{\AA}
Zn$_{0.89}$Mg$_{0.11}$S$_{0.18}$Se$_{0.82}$ barrier. Reflectivity
spectra measured under different illumination intensities are
presented in Fig.~\ref{fig9}. Laser light with $\hbar \omega _L =
$3.5~eV was used for illumination and the high-energy part of the
halogen lamp spectrum was cut by a 420~nm edge filter. Without
laser illumination strong exciton resonances $X_{hh} $ and $X_{lh}
$ dominate the reflectivity spectrum shown by dashed line. Only a
weak $X^{ + }$ resonance is detectable 3.3~meV to low energies
from the $X_{hh} $ one. With increasing illumination power the
$X^{ + }$ resonance vanishes and a new $X^{ - }$ resonance
appears. Its energy distance from the exciton energy is 4.4~meV.
We note here that the same trick has been done in
Ref.~\onlinecite{ref47} for GaAs-based QW's. It is elegant and
very convincing as it allows measuring parameters of $X^{ + }$ and
$X^{ - }$ resonances in the same structure, thus avoiding
technological and growth uncertainties.

Identification of charged exciton resonances were based on their
polarization properties in external magnetic fields (see
Fig.~\ref{fig10}). We conclude from the data of Fig.~\ref{fig9}
that without illumination all donor-electrons from the
modulation-doped layer are either captured by charged surface
states or remain on donors. In this condition the QW contains a
very diluted hole gas with $n_h$=1$\times 10^{10}$~cm$^{-2}$. This
value was determined from the oscillator strength of the $X^{ + }$
transition, which for low carrier densities is linearly
proportional to the carrier concentration (see details in
Refs.~\onlinecite{ref12,ref15}). Laser illumination redistributes
the carrier location in the structure by supplying the QW with
electrons. An increase of the electron density saturates at
$n_e$=6$\times 10^{10}$~cm$^{-2}$ (for detailed behavior see
Fig.~\ref{fig11}), which is still much lower than the
concentration of donors in the modulation doped layer
$n_D$=3$\times 10^{11}$~cm$^{-2}$ evaluated from a technological
calibration. We suggest that the reason is a relatively small
conduction band offset in this structure ($\Delta E_C$=100~meV).
Internal electric fields caused by carrier separation, when only
part of the electrons are removed from the donors, can compensate
for the energy difference between the electron level in the QW and
the barrier donor energy.

\begin{figure}
\includegraphics{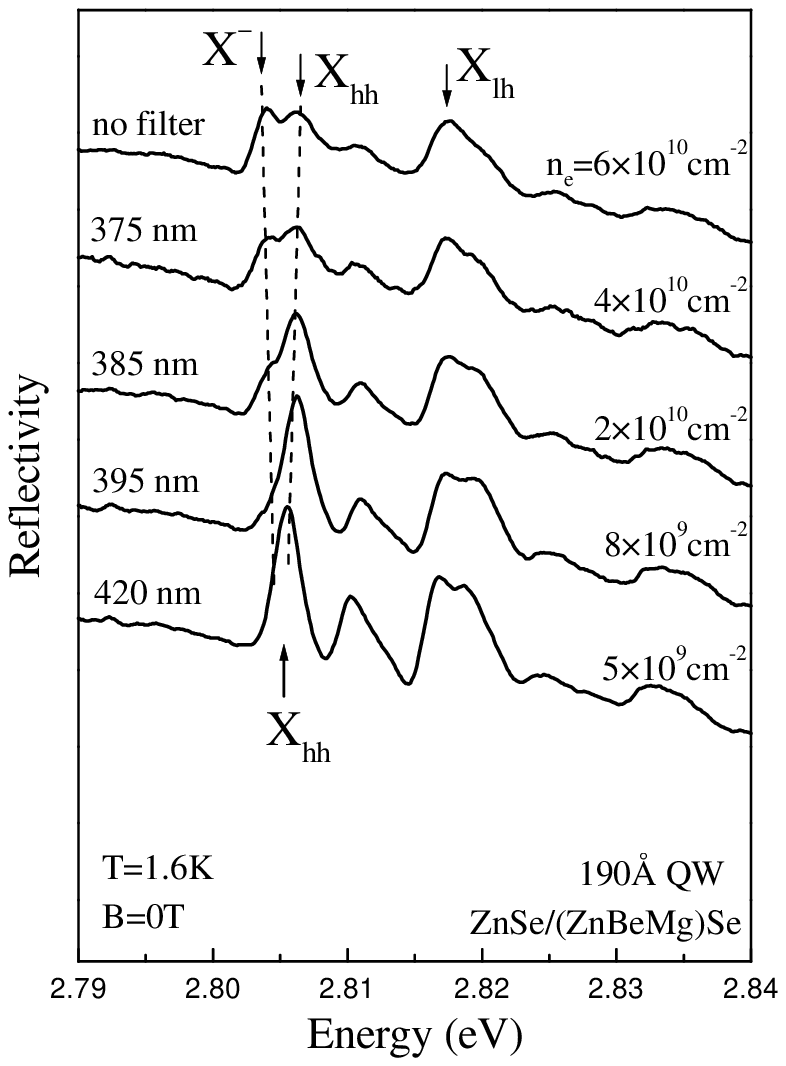}
\caption{\label{fig12} Tuning of electron gas concentration by an
additional above-barrier illumination. The figure presents
reflectivity spectra taken from a 190~{\AA}
ZnSe/Zn$_{0.82}$Be$_{0.08}$Mg$_{0.10}$Se QW as a function of
above-barrier illumination. The illumination is provided by a
white-light source together with longpass optical filters having
different absorption edges. The values of the filter absorption
edge are given in the figure. $B$=0~T, $T$=1.6~K. }
\end{figure}

\begin{figure}
\includegraphics{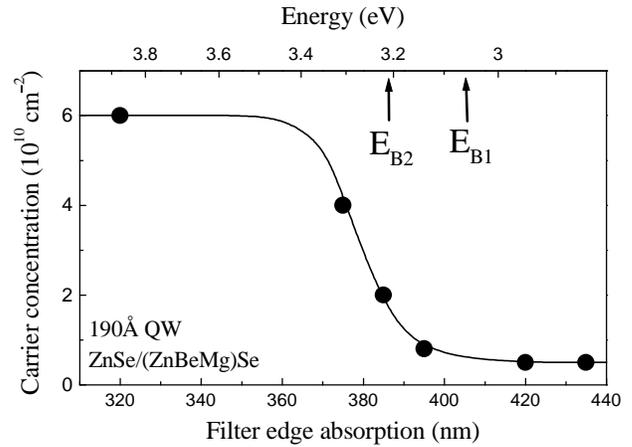}
\caption{\label{fig13} Electron concentration as a function of
wavelength of illuminating light. Symbols are experimental data
from Fig.~\ref{fig12}. Arrows show band gaps of barriers (see
Fig.~\ref{fig1}). Solid line is a guide to the eye. }
\end{figure}

\begin{figure}
\includegraphics{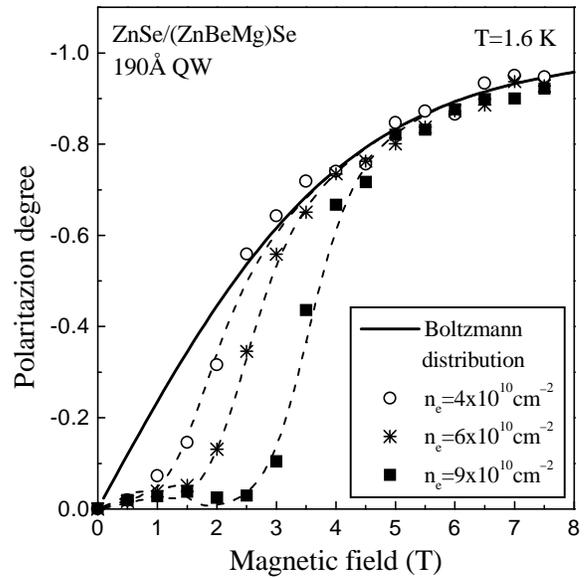}
\caption{\label{fig14} Degree of circular polarization of a
negatively charged exciton {\it vs} magnetic field for different
2DEG densities ($n_e )$ in a 190~{\AA}
ZnSe/Zn$_{0.82}$Be$_{0.08}$Mg$_{0.10}$Se QW. Symbols correspond to
experimental data. Degree of circular polarization for the
nondegenerate 2DEG with g factor $g_e = $+1.12 is shown by a solid
line. Dashed lines represent fittings for the degenerate 2DEG, the
obtained $n_e $ are given in the figure. $T$=1.6~K. }
\end{figure}

In the type A structure cb1172, a 190~{\AA} QW is separated from
the surface by two barriers of different heights. Note that this
structure is nominally undoped. Instead of the laser we use for
illumination the light of the halogen lamp selected by edge
filters. Reflectivity spectra are shown in Fig.~\ref{fig12}. Only
exciton transitions are visible in the spectrum measured with a
420~nm filter, i.e. when photocarriers are excited only in the
ZnSe layer of the 190-{\AA}-thick QW. A threshold-like increase of
the electron density in the QW starts when the energy of
illumination light exceeds the band gap of the highest barrier
(3.21 eV), as clearly seen in Fig.~\ref{fig13}. The electron
density in this structure is varied from 5$\times 10^{9}$ to
9$\times 10^{10}$~cm$^{-2}$. $n_e$ was evaluated from the analysis
of the polarization degree of the trion line. The procedure has
been suggested in Ref.~\onlinecite{ref12} and detailed later in
Ref.~\onlinecite{ref15}. It is based on the fitting of the
magnetic-field-induced polarization of trion resonance in the
frame of the approach accounting for the Fermi-Dirac statistics of
the electron gas. Dashed lines in Fig.~\ref{fig14} show examples
of the fitting. The Fermi energy that is determined from the best
fit of experimental data points is directly linked to the electron
density. From the threshold-like effect of the illumination on the
electron density in the QW shown in Fig.~\ref{fig14} we conclude
that a recharging of surface states (namely a capture of
photo-holes by the surface states) is the main mechanism for the
carrier separation that supplies the QW with free electrons.

\subsection{\label{sub4C} Exciton-trion energy separation. Effect of the Fermi
energy}

\begin{figure}
\includegraphics{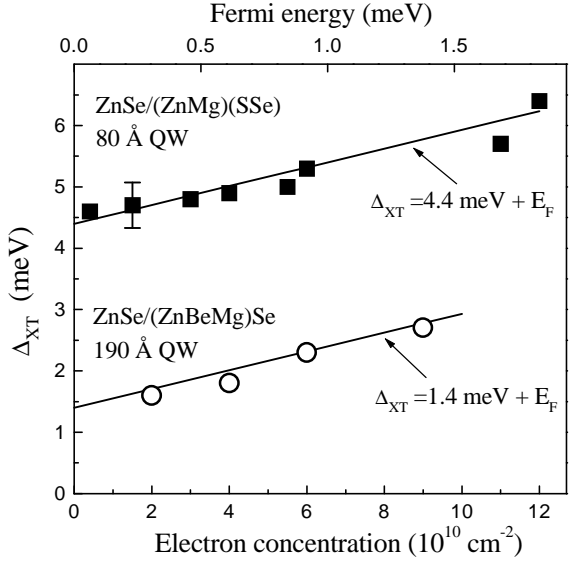}
\caption{\label{fig15} Exciton-trion separation as a function of
2DEG density for a 190~{\AA}
ZnSe/Zn$_{0.82}$Be$_{0.08}$Mg$_{0.10}$Se QW (circles) and an
80~{\AA} ZnSe/Zn$_{0.89}$Mg$_{0.11}$S$_{0.18}$Se$_{0.82}$ QW
(squares). Lines are the sum of a trion binding energy in
corresponding QW and the Fermi energy of a 2DEG. }
\end{figure}

Now we turn our attention to the binding energy of the trions
$E_B^T $, defined as the energy required to dissociate an isolated
trion into a neutral exciton and an electron (for $X^ - $) or a
hole (for $X^{ + }$). In the limit of a very diluted carrier gas,
$E_B^T $ is given by the energy difference between the exciton and
trion lines (i.e. energy separation between bound and unbound
states) $\Delta _{XT} = E_X - E_T $. A deviation from this
``bare'' value of $E_B^T $ takes place with increasing carrier
density. In Fig.~\ref{fig15} we show the exciton-trion separation
$\Delta _{XT} $ as a function of electron density $n_e $ for two
QW's. In the case of an 80~{\AA}
ZnSe/Zn$_{0.89}$Mg$_{0.11}$S$_{0.18}$Se$_{0.82}$ QW (squares) the
electron concentration was varied by modulation doping. For a
190~{\AA} ZnSe/Zn$_{0.82}$Be$_{0.08}$Mg$_{0.10}$Se QW (circles)
the electron density was tuned via additional illumination (see
Sec.~\ref{sub4B}). For both cases the exciton-trion energy
separation increases remarkably with the electron density. Solid
lines in Fig.~\ref{fig15} have a slope of the Fermi energy
increasing with growing electron density $E_F = {\pi \hbar ^2n_e }
\mathord{\left/ {\vphantom {{\pi \hbar ^2n_e } {m_e }}} \right.
\kern-\nulldelimiterspace} {m_e }$. In ZnSe QW's with
$m_e$=0.15$m_0$   $E_F$[meV]=1.53$\times
10^{-11}$~$n_e$[cm$^{-2}$]. Comparing solid lines and data points
in the figure one can establish that the concentration dependence
of the exciton-trion energy separation is approximately given by
the Fermi energy: $\Delta _{XT} = E_B^T + E_F $. This result has
been also reported recently for CdTe-based QW's \cite{ref49}.

Such a behavior of $\Delta _{XT} \left( {n_e } \right)$ does not
correspond to a real increase of the trion binding energy and can
be quantitatively explained in terms of exciton-trion repulsion
due to their mixing \cite{ref17}. This mixing is provided by
mutual transformation of exciton and trion states via exchange of
additional electron. The detailed investigation of mixed
exciton-trion states will be published elsewhere.

To obtain the $E_B^T $ value we extrapolate experimental
dependencies $\Delta _{XT} (n_e )$ to the limit $n_e \to 0$ (see
Fig.~\ref{fig15}) getting the ``bare'' binding energy of trion
$E_B^T $. We performed this procedure for all studied structures
in order to receive information on binding energies of
``isolated'' trions that can be directly compared with theoretical
calculations.

\subsection{\label{sub4D} Binding energy of trions}

\begin{figure}
\includegraphics{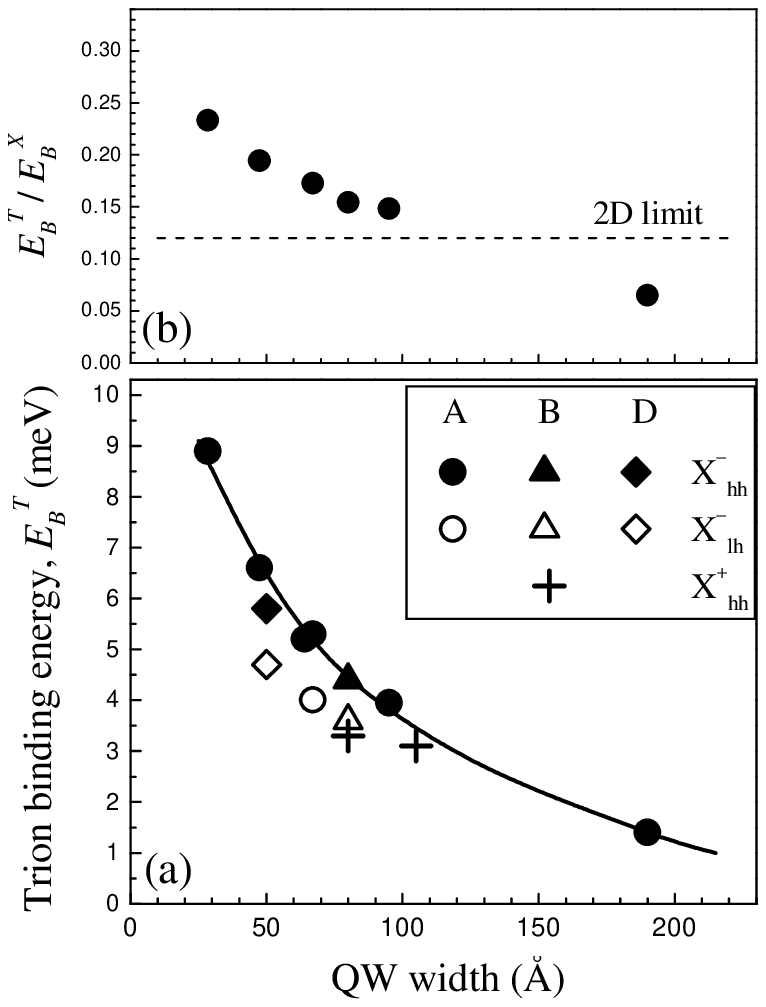}
\caption{\label{fig16} (a) Trion binding energy as a function of
QW width for ZnSe/Zn$_{0.82}$Be$_{0.08}$Mg$_{0.10}$Se (type~A),
ZnSe/Zn$_{0.89}$Mg$_{0.11}$S$_{0.18}$Se$_{0.82}$ (type~B) and
ZnSe/Zn$_{0.96}$Be$_{0.04}$Se (type~D) structures. Solid symbols
show data for negatively charged excitons formed with heavy-hole
($X_{hh}^{ - }$), open symbols are for negatively charged excitons
formed with light-hole ($X_{lh}^{ - }$), and crosses are for
positively charged excitons formed with heavy-hole ($X^{ + }$).
Solid line is interpolation for $X_{hh}^{ - }$ values. The values
of the trion binding energy are also given in Table~\ref{tab4}.\\
(b) Ratio of the $X_{hh}^{ - }$ binding energy to the binding
energy of quasi-two-dimensional exciton taken from
Fig.~\ref{fig5}b. Theoretical value of $E_B^T / E_B^X$=0.12 for a
two-dimensional case is shown by a dashed line \cite{ref66,
ref67}. }
\end{figure}

Binding energies of trions $E_B^T $ determined for the low carrier
density regime are collected in Table~\ref{tab4} and are displayed
in Fig.~\ref{fig16}a as a function of QW width. Solid symbols
correspond to $X_{hh}^{ - }$, open symbols show $X_{lh}^{ - }$ and
crosses are used for $X_{hh}^{ + }$.

\begin{table*}
\caption{\label{tab4}Energetic parameters of excitons and trions
in ZnSe-based QW's.}
\begin{ruledtabular}
\begin{tabular}{cc|c|cccc|ccc}
 \multicolumn{3}{c}{} &\multicolumn{4}{c}{Exciton parameters}&\multicolumn{3}{c}{Trion  binding energy} \\
  & & QW & Exciton &\multicolumn{2}{c}{} & Binding & \multicolumn{3}{c}{(meV)} \\
 SAMPLE & Type & width &energy, $E_X$ &\multicolumn{2}{c}{$E_{1s}-E_{2s}$} & energy & & & \\
 & & ({\AA}) & (eV) & \multicolumn{2}{c}{(meV)}& (meV) & $X_{hh}^{ - }$ & $X_{lh}^{ - }$ & $X_{hh}^{ + }$ \\
 & & & & & Calc.\footnotemark[1] & Calc.\footnotemark[1] & & & \\
 \hline
 cb1175 & A & 29  & 2.8910 &      & 31.0 & 38.2 & 8.9 &     &     \\
 cb1173 & A & 48  & 2.8472 &      & 27.5 & 34.0 & 6.6 &     &     \\
 cb1048 & A & 64  & 2.8277 & 24.0 & 25.0 & 31.2 & 5.2 &     &     \\
 cb1041 & A & 67  & 2.8260 &      & 24.7 & 30.7 & 5.3 & 4.0 &     \\
 cb1174 & A & 67  & 2.8258 &      & 24.7 & 30.7 & 5.3 &     &     \\
 cb1198 & A & 95  & 2.8149 &      & 21.7 & 27.3 & 4.0 &     &     \\
 cb1172 & A & 190 & 2.8057 &      & 16.5 & 21.5 & 1.4 &     &     \\
 \hline
 zq1038 & B & 80  & 2.8182 & 25.0 & 22.7 & 28.5 & 4.4 & 3.6 & 3.3 \\
 zq1113 & B & 105 & 2.8129 &      & 20.6 & 26.1 &     &     & 3.1 \\
 \hline
 zq703  & C & 50  & 2.8260 & 22.5 & 22.7 & 28.6 &     &     &     \\
 \hline
 cb571  & D & 50  & 2.8280 &      & 22.8 & 29.2 & 5.8 & 4.7 &     \\
\end{tabular}
\end{ruledtabular}
\footnotetext[1]{From our calculation.}
\end{table*}

We discuss first the data for the negatively charged excitons
related to heavy-hole excitons. The line in Fig.~\ref{fig16}a is
an interpolation made for $X_{hh}^{ - }$ data points (solid
symbols). The trion binding energy increases strongly from 1.4~meV
in a 190~{\AA} QW up to 8.9~meV in a 29~{\AA} QW. The increase for
$E_B^T $ is 6.4 times while the exciton binding energy increases
only twice (see Fig.~\ref{fig5}b). Stronger sensitivity of the
trion binding energy to confinement conditions is due to the lager
extension of the trion wave function and to the strong effect of
reduction of dimensionality on the trion stability \cite{ref63,
ref69}. Theoretical calculations show that trion states are very
weakly bound in three-dimensional systems, which hinders their
experimental observation in bulk semiconductors. Reduction of
dimensionality from 3D to 2D is a crucial factor for increasing
trion stability, and the trion binding energy grows by a factor of
ten \cite{ref66}. We believe that increase of $E_B^T $ shown in
Fig.~\ref{fig16}a is dominated by localization of carrier wave
functions along the structure growth axis, i.e. by the
increasingly two-dimensional character of the carrier wave
functions. Contribution of the in-plane localization of trions is
minor except perhaps for the very narrow 29~{\AA} QW. This
conclusion is based on comparing the exciton linewidths (see
Fig.~\ref{fig3}), which gives us characteristic energies for the
in-plane exciton localization, with the trion binding energies.
Exciton linewidth is weakly sensitive to the QW width and is below
2~meV for the range 50-190~{\AA}. It increases to 5.3~meV in the
very narrow QW, but even in this case it stays smaller than the
trion binding energy of 8.9~meV.

A detailed comparison of the trion and exciton modifications with
decreasing QW width is given in Fig.~\ref{fig16}b, where the ratio
of the trion and the exciton binding energies ${E_B^T }
\mathord{\left/ {\vphantom {{E_B^T } {E_B^X }}} \right.
\kern-\nulldelimiterspace} {E_B^X }$ is presented. For the
190~{\AA} QW this ratio ${E_B^T } \mathord{\left/ {\vphantom
{{E_B^T } {E_B^X }}} \right. \kern-\nulldelimiterspace} {E_B^X } =
$0.065. It increases linearly with decreasing QW width achieving a
value of 0.235 in the 29~{\AA} QW. Theoretical calculations of
this ratio performed for the two-dimensional limit give a value
${E_B^T } \mathord{\left/ {\vphantom {{E_B^T } {E_B^X }}} \right.
\kern-\nulldelimiterspace} {E_B^X } \approx $0.12, which is rather
insensitive to the ratio of electron and hole effective masses
\cite{ref66,ref67}. The experimental value for the 29~{\AA} QW
exceeds the theoretical limit by a factor of two. We explain this
by the fact that our experimental situation corresponds to the
quasi-2D case rather than to strictly 2D one. This is confirmed by
the moderate increase of the exciton binding energy, which is
twice as large as the bulk Rydberg in narrow QW's and,
respectively, twice as small as the 2D limit of four Rydbergs. A
dimensional transition for a Coulombic state in QW structures is
determined by a ratio of the Bohr radius of the states to the QW
width. Obviously, for trions with larger Bohr radius this
transition will happen in wider QW's than for excitons, whose wave
function is more compact. Thus at a given QW width, excitons and
trions have different degrees of two-dimensionality, which causes
a larger measured value of ${E_B^T } \mathord{\left/ {\vphantom
{{E_B^T } {E_B^X }}} \right. \kern-\nulldelimiterspace} {E_B^X }$
compared with the calculated value for the 2D limit.

It is interesting to note that the strength of confinement
potentials in our structures plays a minor effect on the trion
binding energies. Data points in Fig.~\ref{fig16}a for structures
with different materials with $\Delta E_g $ value varied from 200
to 250~meV follow the same dependence. Only a small deviation from
this dependence was found for the type D structure with $\Delta
E_g $=70~meV. For very shallow 70~{\AA} ZnSe/Zn(S,Se) QW's with
$\Delta E_g $=25-35~meV a trion binding energy of 2.7-2.9~meV has
been reported \cite{ref32}. This is consistent with our data from
Fig.~\ref{fig16}a and evidences that decreasing the electron
confinement leads to smaller binding energies for $X^{ - }$.

We are aware of only one paper where the binding energies of
$X_{hh}^{ - }$ were calculated for ZnSe-based QW's \cite{ref36}.
The quantitative agreement was not satisfactory and authors
suggested that the polaron effect, which in ZnSe QW's could give
an additional 1.3-2.6~meV contribution to the trion binding
energy, should be considered. For our structures we found a
relatively strong dependence of the exciton reduced mass on the QW
width (see Fig.~\ref{fig4}b). We believe that incorporating this
factor into calculations will increase their reliability and
coincidence with experiment. A.Esser has run calculation for
80~{\AA} QW (zq1038) with our new parameters and got a value of
4.2~meV for $X_{hh}^{ - }$ which is in good coincidence with our
experimental value of 4.4~meV even whithout introduction of the
polaronic correction \cite{ref60}.

Binding energies of trions based on the light-hole excitons (open
symbols in Fig.~\ref{fig16}a) are 20-30{\%} smaller than the
$X_{hh}^{ - }$ binding energies. To the best of our knowledge no
detailed investigation of $X_{lh}^{ - }$ states has been reported
so far. Trions associated with light-hole excitons were observed
in PL excitation spectra of GaAs-based QW's \cite{ref64} and in
the reflectivity spectra of monomolecular CdTe islands
\cite{ref65}. In both cases the $X_{lh}^{ - }$ binding energy was
very close to that of $X_{hh}^{ - }$. Numerical calculations
performed for GaAs QW's give, for example, a 40{\%} difference in
favor of $X_{hh}^{ - }$ in a 100~{\AA}
GaAs/Ga$_{0.85}$Al$_{0.15}$As QW \cite{ref63}. However, the model
used in Ref.~\onlinecite{ref63} has not accounted for the
modification of the in-plane effective mass in the valence band
which is essential for the quantitative comparison with
experimental results.

It is interesting that positively charged excitons show binding
energies reduced by about 25{\%} compared with their negatively
charged partners. E.g. in an 80~{\AA} QW of type B (zq1038),
binding energies for negatively- and positively charged excitons
are 4.4 and 3.3~meV, respectively. We are very confident of this
result, as it has been measured in the same structure (see
Figs.~\ref{fig9} and \ref{fig10}) where the type of the free
carriers occupying the QW was reversed by the illumination.
Calculations performed for the 3D and 2D limits give the $X^{ + }$
binding energy larger than the $X^{ - }$ one \cite{ref66,ref67};
this result is explained qualitatively by the heavier effective
mass of holes compared to electrons. However, in the
quasi-two-dimensional case the situation can differ qualitatively.
Recent calculations performed by A.~Esser with parameters of the
structure zq1038 give 4.2 and 4.0~meV binding energies for $X^{-}$
and $X^{ + }$, respectively \cite{ref60}. The smaller binding
energy of the positively charged exciton is explained by the
``effective'' hole-hole Coulomb repulsion to be stronger in this
QW than the electron-electron one. The calculations qualitatively
reproduce experimental trends. Better quantitative agreement for
$X^{ + }$ state is still desired. Note that for GaAs QW's
identical binding energies for $X^{ - }$ and $X^{ + }$ have been
reported \cite{ref47}.

Comprehensive theoretical consideration for the trion binding
energy data collected in Fig.~\ref{fig16}a is still missing. We
hope that these data and the set of exciton parameters given in
the paper will encourage such activity.


\section{\label{sec5}PROPERTIES OF SINGLET TRION STATES}

The singlet state is the ground state of a trion except at very
high magnetic fields where the triplet state gains larger binding
energy. Recently the triplet states have attracted considerable
attention in GaAs-based QW's \cite{ref59,ref61} and in ZnSe-based
structures \cite{ref16}. However, we leave this topic outside the
scope of the present paper and concentrate here on the properties
of the singlet state.

\subsection{\label{sub5A} Magnetic field dependence of binding energy}

\begin{figure}
\includegraphics{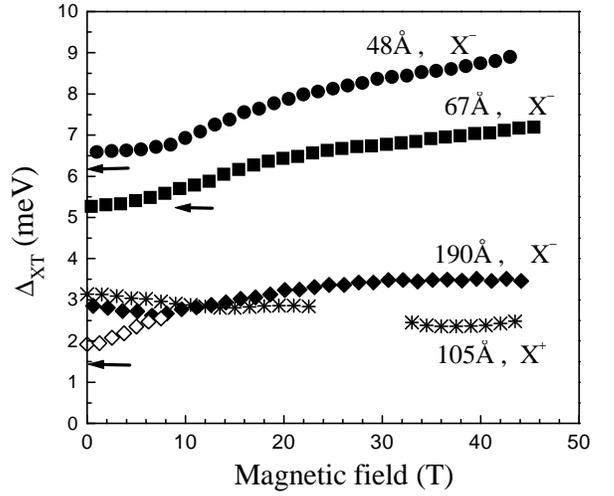}
\caption{\label{fig17} Exciton-trion separation as a function of
magnetic field for QW's of different width. $X^{- }$ was measured
in ZnSe/Zn$_{0.82}$Be$_{0.08}$Mg$_{0.10}$Se QW's: 190~{\AA}
(diamonds), 67~{\AA} (squares), 48~{\AA} (circles). Solid symbols
correspond to the PL excited by laser with energy above barrier
and open symbols are for below-barrier excitation. Arrows indicate
``bare'' trion binding energy. $X^{ + }$ is taken for a 105~{\AA}
ZnSe/Zn$_{0.89}$Mg$_{0.11}$S$_{0.18}$Se$_{0.82}$ QW's (stars).
$T$=1.6 K.}
\end{figure}

Energy distance between exciton and trion PL lines $\Delta _{XT} $
is plotted in Fig.~\ref{fig17} {\it vs} magnetic field strength.
In order to avoid uncertainties caused by spin splittings, data
for the center-of-gravity of exciton- and trion spin doublets are
given. Our task is to study the binding energy of the ``bare''
trions which exhibit no contribution from the Fermi energy. The
regime of the very diluted carrier gas is fulfilled for a 105
{\AA} QW with $X^{ + }$, where $\Delta _{XT} $ at $B$=0~T equals
to $E_B^T $ (see Table~\ref{tab4}). Similar statements can be made
for 48~{\AA} and 67~{\AA} QW's with $X^{ - }$. $E_B^T $ values for
QW's with $X^{ - }$ are shown by arrows. Only in the case of the
190~{\AA} QW was the contribution of the Fermi energy to $\Delta
_{XT} $ considerable for the set of data measured in pulsed
magnetic fields (shown by solid diamonds in Fig.~\ref{fig17}). We
have repeated measurements for this structure in {\it dc} magnetic
fields $B<$8~T keeping the low density of a 2DEG. Results are
given by open diamonds. One can see that the difference between
the two data sets vanishes with increasing magnetic fields and
disappears for $B>$7~T. That means that the contribution of $E_F $
decreases with growing magnetic fields, which can be explained by
an increase of the density of states of the Landau levels. For the
following discussion we consider the $E_B^T (B)$ dependence for a
190~{\AA} QW consisting of open diamonds at low fields and of
solid diamonds at high fields ($B>$8~T).

\begin{figure}
\includegraphics{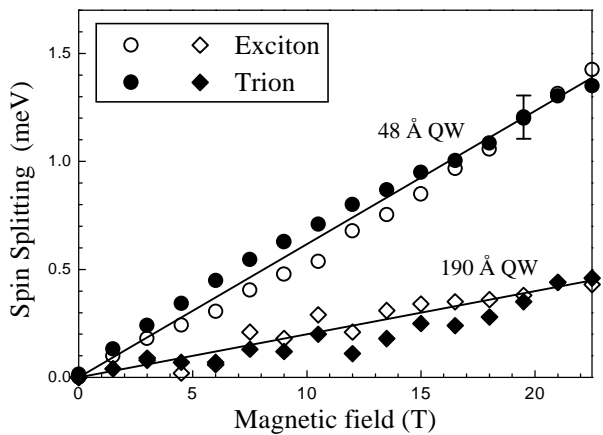}
\caption{\label{fig18} Comparison of exciton and trion Zeeman
splittings in ZnSe/Zn$_{0.82}$Be$_{0.08}$Mg$_{0.10}$Se QW's.
Symbols are experimantal points, full curves are guidelines for
the eye.}
\end{figure}

Binding energies of $X^{ - }$ in all studied QW's show a monotonic
increase with growing magnetic fields and a tendency to saturation
in high fields $B>$25~T. The increase is stronger in wider QW's
with smaller $E_B^T $, e.g. it amounts to 150{\%} in a 190~{\AA}
QW and has only 35{\%} in a 48~{\AA} QW. Being more compact in
narrow QW's the singlet state becomes less sensitive to
compression by external magnetic fields. Qualitatively $E_B^T (B)$
dependencies for $X^{ - }$ from Fig.~\ref{fig17} are consistent
with theoretical predictions for the singlet state
\cite{ref62,ref59,ref16}.

Magnetic field dependence of the positively charged exciton
differs drastically from $X^{ - }$ behavior. Binding energy of
$X^{ + }$ shows no dependence on magnetic fields for $B<$6~T and
decreases by 25{\%} at higher fields (see stars in
Fig.~\ref{fig17}). In the field range 26-32~T the $X^{ + }$
singlet state line shows irregular behavior caused by its crossing
with the triplet state. These results will be published elsewhere
and here, for clarity, we do not show data points for this field
range. Principally different behavior of $X^{ - }$ and $X^{ + }$
states in external magnetic fields has been established first for
GaAs QW's \cite{ref47}. Our results confirm this for ZnSe-based
QW's. We are not aware of theoretical attempts to model $X^{ + }$
behavior in magnetic fields. However, it is clear that the
difference in magnetic field behavior of $X^{ + }$ and $X^{ - }$
binding energies is due to the very different structure of wave
functions of these complexes (see discussion in
Ref.~\onlinecite{ref67}). $X^{ - }$ is constructed of two light
particles (electrons) rotated around one heavy particle (hole).
This complex has one center and magnetic field will localize the
electron wave functions around the hole, thus inducing an increase
of the binding energy. In contrast $X^{ + }$ has two heavy
particles, i.e. two centers, and one light particle moving between
two centers. In this case shrinking of electron wave function by
magnetic fields hinders it from optimal adjustment for two
centers, which results in decreasing binding energy of $X^{ + }$
complex.

\subsection{\label{sub5B}Spin splitting of trions}

In the studied structures, the Zeeman splitting of the trion
singlet state closely follows the behavior of the exciton Zeeman
splitting. Typical examples for 48 and 190~{\AA}
ZnSe/Zn$_{0.82}$Be$_{0.08}$Mg$_{0.10}$Se QW's are given in
Fig.~\ref{fig18}. Deviations between exciton and trion splittings
are inside the error bar of spectral resolution of 0.1~meV. This
result is explained on the basis of a spin structure of trion and
exciton states suggesting that the electron and hole wave
functions in a trion are the same as in a neutral exciton (see
e.g. Refs.~\onlinecite{ref16,ref59}). Indeed the ground state of
the negatively charged exciton exhibits a hole spin splitting as
the two electrons with antiparallel spin orientation are
insensitive to external magnetic fields. However, the Zeeman
splitting of the trion optical transition must also reflect the
Zeeman contribution of the bare electron, which remains after
trion recombination. As a result, the Zeeman splitting of $X^{ -
}$ is given by $g_{hh} - g_e $, which is identical to the exciton
spin splitting. A similar consideration holds for the positively
charged exciton.

Different spin splitting of excitons and trions has been reported
for a 200~{\AA} GaAs-based QW and related to a different mixing of
wave functions in $X^{ - }$ than in $ X $ \cite{ref47}. Also for
ZnSe/Zn(S,Se) QW's with small confinement potential $\Delta E_g =$
35~meV different spin splittings of exciton and negatively changed
exciton have been found \cite{ref32}. We suppose that the small
energy splitting between the heavy-hole and light-hole states in
these structures allows mixing of these states in a trion, that is
resulted in a modification of the hole g factor. In our structures
$\Delta _{lh - hh} $ was relatively large 11-20~meV (see
Table~\ref{tab3}) which prevents the modification of the hole g
factor.

\subsection{\label{sub5C}Oscillator strength of trions}

\begin{figure}
\includegraphics{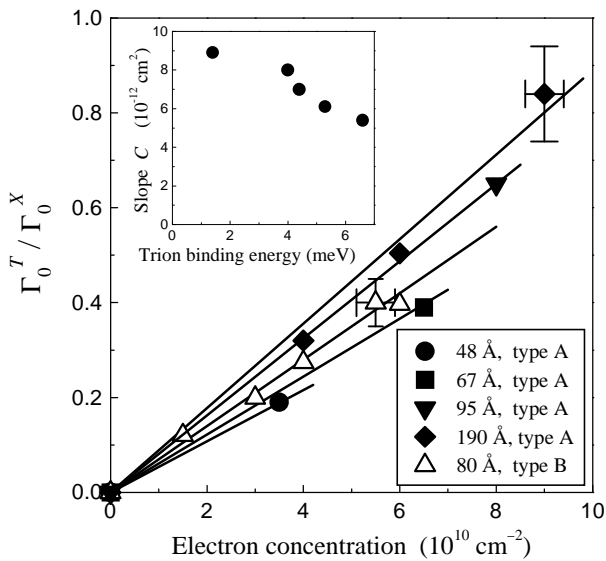}
\caption{\label{fig19} Trion oscillator strength $\Gamma _0^T $
normalized on the exciton one $\Gamma _0^X $ {\it vs} electron
density in different QW's. Lines show linear interpolation of
experimental results. Slope of the dependence $\Gamma _0^T = Cn_e
\Gamma _0^X $ as a function of trion binding energy is given by
the inset.}
\end{figure}

Treatment of resonances in reflectivity spectra allows extracting
the oscillator strength (i.e. radiative damping $\Gamma _0 )$ of
trions and excitons. A detailed study of the trion oscillator
strength in an 80~{\AA}
ZnSe/Zn$_{0.89}$Mg$_{0.11}$S$_{0.18}$Se$_{0.82}$ QW can be found
in Ref.~\onlinecite{ref12}. It has been established experimentally
that $\Gamma _0^T $ grows linearly with electron concentration
$\Gamma _0^T = Cn_e \Gamma _0^X $. The value of $C$ is a very
useful parameter for evaluation of the electron density by optical
method \cite{ref15}.

In Fig.~\ref{fig19} ${\Gamma _0^T } \mathord{\left/ {\vphantom
{{\Gamma _0^T } {\Gamma _0^X }}} \right.
\kern-\nulldelimiterspace} {\Gamma _0^X }$ is plotted as a
function of $n_e $ for QW's of different thickness. Electron
concentration was evaluated from the polarization properties of
$X^{ - }$ in external magnetic fields (see e.g. Fig.~\ref{fig14}).
The value of the slope $C$ increases for wider QW's with smaller
binding energies of trions. This is illustrated in the inset of
Fig.~\ref{fig19} where $C$ is plotted as a function of $E_B^T $.
Trions with the smallest binding energy have the largest extension
of wave function, which covers the largest number of unit cells
and, respectively, gives the largest $\Gamma _0^T $ value. Results
from the inset of Fig.~\ref{fig19} confirm this conclusion. It is
worthwhile to note that that these results also allow
determination of a carrier density in ZnSe-based QW's of various
widths.

\subsection{\label{sub5D} Trion Linewidth}

We discuss here the linewidth of exciton and trion luminescence
lines presented in Fig.~\ref{fig3}. It was found that for QW's
thinner than 100~{\AA}, the trion line is systematically broader
than the exciton one. The difference in linewidths grows up to
60{\%} in a 29~{\AA} QW. At least two physical reasons for that
can be suggested: (i) PL linewidth is contributed by localization
energies for carriers. In the case of excitons it is summed up
from the electron and hole contributions, where the electron plays
a dominant role. In case of trions, two electrons and one hole
participate. Qualitatively it should result in larger broadening,
but the quantitative approach to this problem does not seems to be
very trivial, because it will depend strongly on the choice of a
model for localizing potential. (ii) Another reason is related to
a certain freedom in the energy conservation law in case of the
trion recombination. An electron, which is left after trion
recombination, can have a finite kinetic energy. The energy of
emitted photon will be reduced by this amount. Respectively, the
trion line will exhibit additional broadening due to the electron
kinetic energy.

The first mechanism has a strong dependence on the QW width - its
contribution should increase proportionally with growing
broadening of the exciton line. However, the character of the well
width dependence for the second mechanism is not very obvious for
us.

The second mechanism has been studied theoretically and
experimentally for GaAs-based QW's \cite{ref68}. It was shown that
it has a strong temperature dependence and at $T$=2 K the
additional broadening of the trion line is about 0.04 of the
exciton binding energy. Applying this estimation to our QW's we
get the contribution of the second mechanism of 0.8~meV for a
190~{\AA} QW and of 1.5~meV for a 29~{\AA} QW. In the narrow QW,
exciton and trion linewidths are 4 and 6.5~meV, respectively, i.e.
they differ by 2.5~meV. From this we suggest that both mechanisms
have comparable contribution to the broadening of the trion
emission line. Further experiments including the careful analysis
of the temperature dependencies of the trion linewidth are
required to separate the role of two mechanisms.


\section{\label{sec6} CONCLUSIONS}

Negatively and positively charged excitons in ZnSe-based QW's were
investigated in structures with various QW widths and free carrier
densities. The binding energy of $X^{ - }$ shows a strong
dependence on the QW width, increasing from 1.4 to 8.9~meV as the
well width decreases from 190 to 29~{\AA}. This variation is 6.4
times while the neutral exciton binding energy increases only
twice. The binding energy of $X^{ + }$ is 25{\%} smaller than that
of $X^{ - }$. This observation is in qualitative agreement with
model calculations and is explained by stronger ``effective''
Coulomb repulsion in case of hole-hole interaction compared with
electron-electron interaction. Qualitatively different behavior
for $X^{ - }$ and $X^{ + }$ is found in external magnetic fields.
$X^{ - }$ increases its binding energy depending on the QW width
by 35-150{\%}, while in contrast $X^{ + }$ shows a decrease of its
binding energy by 25{\%}. A detailed set of exciton parameters for
the studied structures is collected in the paper. We hope that
this will encourage theoretical efforts for better understanding
the energy- and spin structure of trions.


\begin{acknowledgments}

We acknowledge stimulating discussions with A.~B.~Dzyubenko and
R.~A.~Suris. We are thankful to A.~Esser for allowing us to use in
this paper his unpublished results on calculation of trion binding
energies. The work was supported in part by the Deutsche
Forschungsgemeinschaft through Sonderforschungbereich 410 and
grant Nos.~Os98/6, 436 RUS~113/557 and He-1939/16-1, as well as by
grants of the Russian Foundation for Basic Research (grants
Nos.~00-02-04020 and 01-02-04010)

We feel deep sorrow at the death of our colleague W.~Faschinger,
who was an outstanding scientists and very kind person.

\end{acknowledgments}


\bibliography{paper}

\end{document}